\documentclass[fleqn,usenatbib]{mnras}

\usepackage{graphicx, color}

\usepackage{newtxtext,newtxmath}

\usepackage[T1]{fontenc}
\usepackage{ae,aecompl}

\def\kcdm{k_{\rm CDM}}
\def\khdm{k_{\rm HDM}}
\def\kM{k_{\rm M}}
\def\ktau{k_{\rm \tau}}
\def\nbar{n_B}
\def\npol{n}
\def\KI{K_{I}}
\def\KII{K_{II}}
\def\Mcrit{M_{\rm crit}}
\def\rhocrit{\rho_{\rm crit}}
\def\xcrit{x_{\rm crit}}

\title[Supermassive Stars at the Onset of Collapse]{Maximally Rotating Supermassive Stars at the Onset of Collapse: The Perturbative Effects of Gas Pressure, Magnetic Fields, Dark Matter and Dark Energy}
\author[Butler, Lima, Baumgarte \& Shapiro]{Satya P.~Butler,$^{1}$ 
Alicia R.~Lima,$^{1}$ 
Thomas W.~Baumgarte,$^{1}$ 
and Stuart L.~Shapiro$^{2,3}$ \\
 $^{1}$Department of Physics and Astronomy, Bowdoin College, Brunswick, ME 04011\\
 $^{2}$Department of Physics, University of Illinois at Urbana-Champaign, Urbana, IL 61801\\
 $^{3}$Department of Astronomy and NCSA, University of Illinois at Urbana-Champaign, Urbana, IL 61801
}

\date{Accepted XXX. Received YYY; in original form ZZZ}

\pubyear{2018}

\begin{document}
\label{firstpage}
\pagerange{\pageref{firstpage}--\pageref{lastpage}}
\maketitle

%
\begin{abstract}
The discovery of quasars at increasingly large cosmological redshifts may favor ``direct collapse" as the most promising evolutionary route to the formation of supermassive black holes.  In this scenario, supermassive black holes form when their progenitors -- supermassive stars -- become unstable to gravitational collapse.   For uniformly rotating stars supported by pure radiation pressure and spinning at the mass-shedding limit, the critical configuration at the onset of collapse is characterized by universal values of the dimensionless spin and radius parameters $J/M^2$ and $R/M$, independent of mass $M$.  We consider perturbative effects of gas pressure, magnetic fields, dark matter and dark energy on these parameters, and thereby determine the domain of validity of this universality.  We obtain leading-order corrections for the critical parameters and establish their scaling with the relevant physical parameters.  We compare two different approaches to approximate the effects of gas pressure, which plays the most important role, find identical results for the above dimensionless parameters, and also find good agreement with recent numerical results.
\end{abstract}
%
\begin{keywords}
black hole physics -- stars: Population III -- equation of state 
\end{keywords}

%
\section{Introduction}
\label{sec:intro}
%

Quasars and active galactic nuclei, believed to be powered by accreting supermassive black holes (SMBHs), are observed out to large cosmological distances \citep[see, e.g.,][]{Fan06,Fanetal06}.  \citet{Banetal17} recently announced the discovery of the most distant quasar ever observed, J1342+0928,  at a redshift of $z \simeq 7.5$, and powered by a SMBH with mass of approximately $7.8 \times 10^8 M_\odot$.   The previous distance-record holder was J1120-0641, at a redshift of $z \simeq 7.1$ and with a black-hole mass of approximately $2.0 \times 10^9 M_{\odot}$ \citep{Moretal11}.   Another remarkable object is the ultra-luminous quasar J0100+2802 discovered by \citet{Wuetal15}, at a redshift of $z = 6.3$ and with a mass of about $1.2 \times 10^{10} M_\odot$.  The detection of these objects poses an important astrophysical problem that has attracted significant attention \citep[see, e.g.,][for reviews]{Sha04,Hai13,LatF16,SmiBL17}: how could such massive black holes form in such a short time after the big bang? 

One evolutionary scenario for the formation of SMBH invokes first-generation -- i.e.~Population III (Pop III) -- stars \citep{MadR01,HegW02}.  These stars could collapse
to form seed black holes \citep{FryWH01,HegFWLH03} at large cosmological redshift, which, conceivably, could then grow through accretion and/or mergers to form SMBHs.  For a given efficiency $\epsilon$ of the conversion of matter to radiation, growth by accretion is usually limited by the Eddington luminosity \citep{Sha05,PacVF15}.    It has been suggested that episodic super-Eddington accretion may speed up the growth of seed black holes \citep{VolR05,VolSD15,LupHDFMM16,SakIH16}.  On the other hand, the effects of photoionization and heating appear to reduce accretion to only a fraction of the Eddington limit (see \citet{AlvWA09,MilBCO09}, see also \citet{WhaF12} for the effect of natal kicks on the accretion rate).   Black holes can also grow through mergers \citep[see, e.g.][]{VolHM03,VolMH03,Sha05,TanH09,Tan14}, even though some limits on growth by mergers may be imposed by recoil speeds \citep{Hai04}.

Given these constraints it is difficult to see how seed black holes with masses of Pop III stars, about 100 $M_{\odot}$, could grow to the masses of SMBHs by $z \simeq 7$.   In fact, \citet{Banetal17} argue that the existence of the objects J1342+0928, J1120-0641 and J0100+2802 ``is at odds with early black hole formation models that do not involve either massive ($\gtrsim 10^4 M_{\odot}$) seeds or episodes of hyper-Eddington accretion" (see also their Fig.~2).   These considerations suggest the direct collapse of objects with masses of $M \gtrsim 10^{4-5} M_{\odot}$ as a plausible alternative scenario for the formation of SMBHs \citep[e.g.][]{Ree84,LoeR94,OhH02,BroL03,KouBD04,Sha04,LodN06,BegVR06,RegH09b,Beg10,AgaKJNDVL12,JohWLH13}.

The ``direct-collapse" scenario assumes that the progenitor object, which we will refer to as a supermassive star (SMS), is able to avoid fragmentation.   Physical processes that help suppress fragmentation are turbulence \citep[e.g.][]{WisTA08,LatSSN13,MayFBQRSW15} and a Lyman-Werner radiation background \citep[see][and references therein]{BroL03,RegH09a,VisHB14}.  The Lyman-Werner radiation dissociates molecular hydrogen, the most efficient coolant in metal-free halos, which otherwise would allow the halo to cool to such low temperatures that its Jeans mass would become small compared to its mass itself, thereby leading to fragmentation \citep[see also][]{LiKML03}.  In fact, the recent discovery of the strong Lyman-$\alpha$ emitter CR7 at $z = 6.6$ \citep{Sobetal15} has been interpreted as observational evidence of the direct collapse scenario (see \citet{SmiBL16,Haretal16,AgaJZLBNK16}; also see \citet{Bowetal17,AgaJKPRKO17} for a discussion of the role of metals in this object).   

The formation, evolution, stability and collapse of SMSs have been the subject of numerous studies over several decades (see, e.g., \citet{Ibe63,HoyF63,Cha64,BisZN67,Wag69,AppF72,BegR78,FulWW86} for some early references, as well as \citet{ZelN71} and \citet[][hereafter ST]{ShaT83} for textbook treatments).   While the existence of SMSs had been considered somewhat hypothetical for many years, and while many questions concerning the formation of SMSs remain open (see, e.g., \citet{SchPFGL13,HosYIOY13,SakHYY15,UmeHOY16,WooHWHK17,HaeWKHW18a,HaeWKHW18b} for recent studies; see also \citet{SmiBL17}), the discovery of objects like J1342+0928, J1120-0641 and J0100+2802 and CR7 suggests that they not only exist, but that they played a key role in the formation of SMBHs.

These considerations have motivated us to revisit an idealized evolutionary scenario for rotating SMSs that two of us proposed previously \citep[see][hereafter Paper I]{BauS99b}.   Specifically, we assumed that SMSs are dominated by radiation pressure, and that turbulent viscosity produced by magnetic fields transports angular momentum sufficiently efficiently to maintain uniform rotation \citep{BisZN67,Wag69}.  We further assumed that SMSs, after initially formed, cool and contract, leading to a spin-up.  Given our assumption of uniform rotation, the star will ultimately reach mass-shedding, i.e. the Kepler limit. While it has been suggested that SMSs formed from accretion only rotate at a fraction of the mass-shedding limit \citep[e.g.][]{MaeM00,HaeWKHW18a}, such stars may still reach mass-shedding during subsequent cooling and contraction. It is also possible that alternative formation scenarios lead to SMSs that rotate rapidly, arriving at mass-shedding either initially or during subsequent cooling and contraction.  Once having reached the Kepler limit, the SMS evolves along the mass-shedding limit \citep[at a rate computed in][]{Baus99a} until it reaches an onset of radial instability, triggering collapse to a black hole.   The onset of instability is a consequence of the interplay between the stabilizing effects of rotation and the destabilizing effects of relativistic gravitation.  Remarkably, the critical configuration marking the onset of instability is characterized by unique values of the dimensionless parameters $J/M^2$ and $R_p/M$, where $J$ is the total angular momentum, $M$ the mass, and $R_p$ the polar radius.  The equatorial radius is given by $R_{\rm eq} = 3 R_p/2$.

The uniqueness of these parameters has profound implications for the subsequent collapse to black holes, because it then has to follow a unique evolutionary track as well.   A number of authors have studied this rotating collapse numerically in the context of general relativity, starting with \citet{ShiS02} (see also \citet{ShaS02} for a related analytical treatment).  The collapse results in a spinning black hole with mass $M_{\rm BH}/M \simeq 0.9$ and angular momentum $J_{\rm BH}/M_{\rm BH} \simeq 0.7$, surrounded by a disk with mass $M_{\rm disk}/M \simeq 0.1$.   The universal evolutionary track also emits a universal gravitational wave signal \citep{ShiSUU16,SunPRS17}, which might serve as a ``standard candle" for future space-based gravitational wave detectors.  Observations of this ``signature" gravitational wave signal would firmly establish SMSs as the progenitors of SMBHs.   

The original simulations of \citet{ShiS02}, who modeled the star as an ideal radiation fluid, were followed up by several other studies in general relativity that also allowed for gas pressure \citep{ShiSUU16}, nuclear reactions \citep{MonJM12,UchSYSU17}, as well as magnetic fields \citep{LiuSS07,SunPRS17}.  The latter demonstrate that this collapse may result in the launch of a jet and power ultra-long gamma-ray bursts, and suggest that collapsing SMSs are promising multimessenger sources for coincident gravitational and electromagnetic radiation.

Given the importance of the dimensionless parameters describing the critical configurations, it is of interest to evaluate to what degree their universality depends on the assumptions made in their derivation -- in particular the assumption of a pure radiation fluid -- and to establish the domain of validity of this universality.  In this paper we develop an analytic perturbational approach to study the effects of gas pressure, magnetic fields, dark matter and dark energy on the critical configuration of maximally rotating SMSs and its dimensionless parameters; our results are summarized in eqs.~(\ref{JoverM2_summary}) and (\ref{RoverM_summary}) below.  For the astrophysical scenarios that we consider most realistic, gas pressure plays the most important perturbative role by far.  Accordingly, most of our paper focusses on gas pressure.  We compare two different approximations that have been adopted to account for gas pressure (see Sections \ref{sec:thermo:appI} and \ref{sec:thermo:appII} below), and calibrate these two different approaches for nonrotating stars (Section \ref{sec:nonrotSMS}) before applying them to maximally rotating SMSs (Section \ref{sec:gas}).

Our perturbative approach builds on an analytical calculation employing a simple energy functional and variational principle that we adopted in Paper I to identify the critical configuration (see Sections~\ref{sec:rotSMS:review} and \ref{sec:rotSMS:background} below).  In a completely independent approach, we also constructed numerically fully relativistic equilibrium models of rotating stars in Paper I.  We found good agreement in the parameters characterizing the critical configuration between the two approaches (see Table 2 in Paper I as well as Table \ref{table:crit} below); typically these parameters agree to within 10\% or better. \citet[][hereafter SUS]{ShiUS16} recently generalized these numerical results, adopting what we refer to below as Approach II to approximate the effects of gas pressure.  Our analytic calculation presented here complements those numerical results: we find good agreement between our analytical results and the numerical results of SUS (see, e.g., Fig.~\ref{fig:compare} below), and we also extend our analytical treatment to account for the effects of magnetic fields, dark matter and dark energy (Section \ref{sec:others}).

This paper is organized as follows.  In Section \ref{sec:thermo} we review some thermodynamic relations.  In particular, we discuss in Sections \ref{sec:thermo:appI} and \ref{sec:thermo:appII} the two different approaches to modeling gas pressure as a small perturbation to radiation pressure.  In Section \ref{sec:nonrotSMS} we compare these two approaches for nonrotating SMSs.  We construct numerical models in Section \ref{sec:nonrotSMS:TOV}, and use these to calibrate an analytical model calculation in Section \ref{sec:nonrotSMS:ana}.   In Section \ref{sec:rotSMS} we extend the analytical model calculations to rotating SMSs.  We review the general setup from Paper I and compute the unperturbed background solution in Sections \ref{sec:rotSMS:review} and \ref{sec:rotSMS:background}, and develop a general framework for perturbations in Section \ref{sec:rotSMS:perturb}.  In Section \ref{sec:gas} we return to the effects of gas pressure.  We adopt Approaches I and II in Sections \ref{sec:gas:appI} and \ref{sec:gas:appII}, compare with the numerical results of SUS in Section \ref{sec:gas:comp}, and provide estimates for physical parameters in Section \ref{sec:gas:results}.  In Section \ref{sec:others} we apply our perturbative approach to estimate the effects of magnetic fields, a dark-matter halo, and dark energy on the critical configuration of SMSs that are uniformly rotating at the mass-shedding limit.   We provide a brief summary in Section \ref{sec:summary}.  

Unless noted otherwise we adopt geometrized units with $c = G = 1$.

%
\section{Thermodynamic Preliminaries}
\label{sec:thermo}
%

In this Section we review some thermodynamic relations, following the treatment in several textbooks (e.g.~ST as well as \citet{Cla83} and \citet{KipWW12}).  While most of these relations are well known, we list them here for an easier comparison of two different approaches to modeling the effect of gas pressure on SMSs, which we introduce in Sections \ref{sec:thermo:appI} and \ref{sec:thermo:appII}.

%
\subsection{Radiation Pressure}
\label{sec:thermo:rad}
%

For a pure thermal radiation fluid the pressure $P_r$ is given by
\begin{equation} \label{P_rad}
P_r = \frac{1}{3} a T^4,
\end{equation}
and the internal energy density $\epsilon_r$ by
\begin{equation} \label{eps_rad}
\epsilon_r = a T^4,
\end{equation}
where $T$ is the temperature and $a$ the radiation constant
\begin{equation}\label{radiation_constant}
a = \frac{8 \pi^5 k_B^4}{15 c^3 h^3}
\end{equation}
(with $c = 1$ in geometrized units).  Here $k_B$ is the Boltzmann constant and $h$ is Planck's constant.  From the first law of thermodynamics
\begin{equation} \label{first_law}
T ds = d \Big( \frac{\epsilon}{\nbar} \Big) + P \, d \Big( \frac{1}{\nbar} \Big),
\end{equation}
where $\nbar$ is the baryon number density, we find that the photon entropy per baryon is
\begin{equation} \label{ent_rad}
s_r = \frac{4a}{3} \, \frac{T^3}{\nbar}.
\end{equation}
Combining eqs.~(\ref{P_rad}) and (\ref{ent_rad}) we can write the pressure as
\begin{equation}
P_r = K_r \rho_0^{\Gamma}
\end{equation}
with $\Gamma = 1 + 1/\npol = 4/3$.  Here $\rho_0 = \nbar m_B$ is the rest-mass density, with $m_B$ the baryon rest mass, and we have defined
\begin{equation} \label{K}
K_r \equiv \frac{a}{3} \left( \frac{3 s_r}{4 m_B a}  \right)^{4/3}.
\end{equation} 
Evidently, stars dominated by radiation pressure behave as $\npol = 3$ polytropes for constant entropy.

%
\subsection{Gas Pressure}
\label{sec:thermo:gas}
%

If gas pressure cannot be neglected, the total pressure and internal energy density are given by
\begin{equation} \label{P_tot}
P = P_r + P_g
\end{equation}
and 
\begin{equation} \label{eps_tot}
\epsilon = \epsilon_r + \epsilon_g,
\end{equation}
where 
\begin{equation} \label{P_gas}
P_g = Y_T \nbar k_B T
\end{equation}
is the gas pressure and
\begin{equation} \label{eps_gas}
\epsilon_g = \frac{3}{2} Y_T \nbar k_B T
\end{equation}
the internal energy density of the plasma.  Here $Y_T$ is the number of particles per baryon.  For simplicity we will assume a fully ionized hydrogen gas in the following, in which case $Y_T = 2$. 
The total entropy per baryon is then given by
\begin{equation} \label{ent_tot}
s = s_r + s_g,
\end{equation}
where $s_g$ is the gas entropy
\begin{equation} \label{ent_gas}
\frac{s_g}{k_B} =  \ln \left( 4 \frac{m_e^{3/2} m_B^{3/2}}{\nbar^2} \left( \frac{k_B T}{2 \pi \hbar^2} \right)^{3} \right) + 5 
= \ln \frac{T^3}{\rho_0^2} + \frac{s_0}{k_B}
\end{equation}
with
\begin{equation} \label{s0}
\frac{s_0}{k_B} 
= 3 \ln \left( \frac{k_B}{2 \pi \hbar^2} \right) + \frac{3}{2} \ln m_e + \frac{7}{2} \ln m_B + 2 \ln 2 + 5
\end{equation}
(see eq.~(17.3.4) in ST; hereafter (ST.17.3.4)).  Here $m_e$ is the electron mass.

%
\subsection{Eddington's Argument}
\label{sec:thermo:eddington}
%

The Eddington standard model (\citet{Edd18}; see also \citet{Cha39,Cla83,KipWW12}, as well as many other references) is based on the observation that, if the ratio\footnote{In many texts the ratio $\beta$ is alternatively defined to refer to the ratio $P_g / P$ rather than $P_g/P_r$.}
\begin{equation} \label{beta}
\beta \equiv \frac{P_g}{P_r} = \frac{8 k_B}{s_r}
\end{equation} 
is constant throughout a given star, then the star again behaves like an $\npol = 3$ polytrope, i.e.~the total pressure again satisfies a polytropic relation 
\begin{equation} \label{P_pol_Edd}
P = K_{E} \rho_0^\Gamma
\end{equation}
with $\Gamma = 4/3$.  This can be seen by writing
\begin{equation}
P = (1 + \beta) P_r = \frac{1 + \beta}{3} a T^4.
\end{equation}
This relation can now be used to eliminate $T$ in terms of $P$ in both terms on the right-hand side of (\ref{P_tot}).  Solving the result for $P$ yields (\ref{P_pol_Edd}) with
\begin{equation} \label{K_edd}
K_{E} = (1 + \beta) \frac{a}{3} \left( \frac{3 s_r}{4 m_B a} \right)^{4/3}  = (1 + \beta) K_r.
\end{equation}
As expected, $K_{E}$ reduces to $K_r$ in the limit $\beta \rightarrow 0$.  

Under certain circumstances it is reasonable to approximate $\beta$ as constant.  SMSs are expected to be convective, however \citep[see, e.g.][]{LoeR94}, which makes it more realistic to assume the {\em total} entropy $s$, rather than the radiation entropy $s_r$ in (\ref{beta}), to be constant inside a given star.  This assumption forms the basis of our first approach to modeling the effects of gas pressure on supermassive stars

%
\subsection{Effects of gas pressure: Approach I}
\label{sec:thermo:appI}
%

In what we call ``Approach I" we  follow Section 17.3 in ST to treat the gas pressure as a small perturbation to the radiation pressure at constant total entropy $s$.  Specifically, we assume that $s_g \ll s_r$ in (\ref{ent_tot}), so that, to zeroth order, the temperature is given by (\ref{ent_rad}) with $s_r$ replaced by $s$.  The leading-order correction to the temperature in terms of $s$ can then be computed by inserting (\ref{ent_rad}) and (\ref{ent_gas}) into (\ref{ent_tot}); this results in
\begin{equation} \label{T_1}
T \simeq \left( \frac{3 s \rho_0}{4 m_B a} \right)^{1/3} \left( 1 - \frac{s_0}{3s} - \frac{k_B}{3 s} \ln \frac{3 s}{4 m_B a \rho_0} \right)
\end{equation}
(see (ST.17.3.9)).  Inserting this into (\ref{eps_tot}) then yields
\begin{equation}
\epsilon \simeq \KI \rho_0^{4/3} \left( 3 + \bar \lambda + \bar \mu \ln \rho_0 \right).
\end{equation}
Here we defined
\begin{equation} \label{K_I}
\KI \equiv \frac{a}{3} \left( \frac{3 s}{4 m_B a}  \right)^{4/3},
\end{equation}
and the coefficients $\bar \lambda$ and $\bar \mu$ are given by
\begin{equation} \label{lambda}
\bar \lambda =  - \frac{4 s_0}{s} + \frac{12 k_B}{s} - \frac{4 k_B}{s} \ln \frac{3s}{4 m_B a}
\end{equation}
and 
\begin{equation} \label{mu}
\bar \mu = \frac{4 k_B}{s}.
\end{equation}
ST list the related coefficients $\lambda = \KI \bar \lambda$ and $\mu = \KI \bar \mu$ in their eqs.~(ST.17.3.11) and (ST.17.3.12).

Since $\bar \lambda$ and $\bar \mu$ describe leading-order corrections, we may replace $s$ in (\ref{lambda}) and (\ref{mu}) with $s_r$, which, using (\ref{beta}), can then be expressed in terms of $\beta$.  Also inserting (\ref{s0}) and (\ref{radiation_constant}) into (\ref{lambda}) and (\ref{mu}) we obtain
\begin{equation} \label{eps_appI}
\epsilon \simeq \KI \rho_0^{4/3} \left( 3 - \beta \left(1 - \frac{5}{2} \ln \beta - \frac{1}{2} \ln \left( \KI^3 \rho_0 \right)  + \frac{1}{2} \ln \eta \right) \right),
\end{equation}
where we have used the abbreviation
\begin{equation} \label{eta}
\eta = \frac{2^4 3^4 5^2}{\pi^7} \left( \frac{m_e}{m_B} \right)^{3/2} \simeq 1.367 \times 10^{-4}.
\end{equation}
We note that, in geometrized units, the combination $\KI^3 \rho_0$ is dimensionless (see Section \ref{sec:nonrotSMS:TOV} below), so that all arguments of logarithms in (\ref{eps_appI}) are dimensionless numbers.

For a given value of the entropy $s$, the internal energy density $\epsilon$ is given as a function of the rest-mass density $\rho_0$ by (\ref{eps_appI}).  The pressure $P$ can be computed by observing that
\begin{equation} \label{eps_I}
\epsilon =  3 P_r + \frac{3}{2} P_g = 3 \left( 1 + \frac{\beta}{2} \right) P_r
\end{equation}
and therefore
\begin{equation} \label{P_I}
P = \left( 1 + \beta \right) P_r = \frac{1}{3} \, \frac{1 + \beta}{1 + \beta/2} \, \epsilon.
\end{equation} 
Since $P_g$ is again considered a small correction to $P_r$, we can again approximate $\beta$ as given by (\ref{beta}) with $s_r$ replaced by $s$ in these leading-order correction terms.  For $\beta \rightarrow 0$ we evidently recover the radiation-fluid expressions of Section \ref{sec:thermo:rad}.  We note that, for nonzero $\beta$, Approach I does not assume the equation of state (EOS) to be of polytropic form.

%
\subsection{Effects of gas pressure: Approach II}
\label{sec:thermo:appII}
%

An alternative approach to approximating the effects of gas pressure is based on the observation that, in the presence of both radiation and gas pressure, the adiabatic exponent is given by
\begin{equation} \label{Gamma1}
\Gamma_1 \equiv \left( \frac{d \ln P}{d \ln \rho_0} \right)_s = \frac{4}{3} +  \frac{\beta (4 + \beta)}{3(1 + \beta)(8 + \beta)} 
\simeq \frac{4}{3} + \frac{\beta}{6},
\end{equation}
where $\beta$ is again given by (\ref{beta}) (see, e.g., \citet{Edd18b,Cha39,BonAC84}; see also Problem 17.3 in ST and Problem 2.26 in \citet{Cla83}).  This suggests that we may approximate the EOS as
\begin{equation} \label{P_II}
P = \KII \, \rho_0^{\Gamma_1}.
\end{equation}
Using the arguments of Section \ref{sec:thermo:eddington} we find
\begin{equation} \label{K_II}
\KII = K_E \, \rho_0^{-\beta/6}
\end{equation}
with $K_E$ given by (\ref{K_edd}) (compare, for example, eq.~(7) in SUS).   Evidently, $\KII$ is only approximately constant for small but nonzero $\beta$.  Approximating $\KII$ as constant, however, the internal energy density is given by
\begin{equation} \label{eps_II}
\epsilon = n_1 P
\end{equation}
where
\begin{equation} \label{n_1}
n_1 = \frac{1}{\Gamma_1 - 1} = \frac{3}{1 + \beta/2}
\end{equation} 
is the approximate polytropic index.

%
\section{Effects of gas pressure on nonrotating SMSs}
\label{sec:nonrotSMS}
%

In this Section we adopt both Approaches I and II to consider the effects of gas pressure on nonrotating stars.  While most of these results can be found in the literature, nowhere are the two approaches carefully distinguished or compared.  Hence we include this Section in order to (a) compare predictions from Approaches I and II, and (b) calibrate an analytical model that we will apply to rotating stars in Section~\ref{sec:gas}.

%
\subsection{Numerical Results}
\label{sec:nonrotSMS:TOV}
%

The structure of a relativistic, spherically symmetric SMS is governed by the Tolman-Oppenheimer-Volkoff (TOV) equations
\begin{equation} \label{TOV}
\begin{array}{rcl} 
\displaystyle \frac{dm}{dr} & = & \displaystyle 4 \pi (\rho_0 + \epsilon) r^2 \\[2mm]
\displaystyle \frac{dP}{dr} & = & \displaystyle - (\rho_0 + \epsilon + P) \frac{m + 4 \pi P r^3}{r^2 (1 - 2m/r)},
\end{array}
\end{equation}
where $m(r)$ is the mass enclosed within a radius $r$, and $M = m(R)$ is the total mass, with $R$ being the stellar radius \citep[see][]{OppV39,Tol39}.

In geometrized units, $\epsilon$ and $\rho_0$ have the same units of inverse length squared.  From eqs.~(\ref{eps_I}) and (\ref{eps_II}) we therefore see that $K^{n/2}$ must have units of length, where $K = \KI$ and $n = 3$ in Approach I, and $K = \KII$ and $n = n_1$ in Approach II (note, though, that the EOS is not assumed to be of polytropic form in Approach I).  In either approach we may therefore introduce dimensionless quantities, for example $\bar m = K^{-n/2} m$ and $\bar \rho_0 = K^{n} \rho_0$.  When written in terms of these dimensionless quantities, the TOV equations (\ref{TOV}), together with (\ref{eps_I}) and (\ref{P_I}) in Approach I and (\ref{P_II}) and (\ref{eps_II}) in Approach II, become independent of the constants $K$.

\begin{figure}
\begin{center}
\includegraphics[width = 3in]{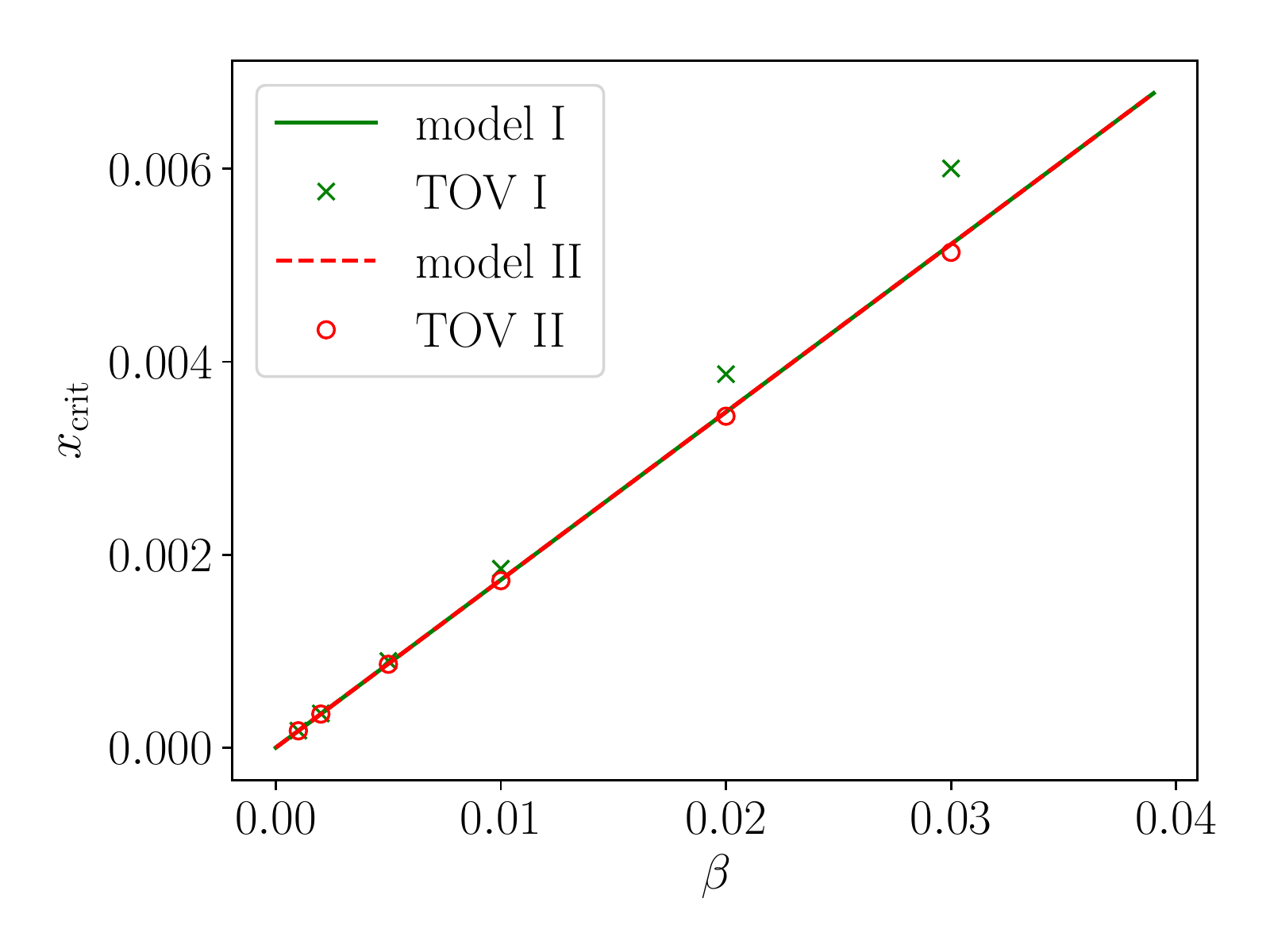}
\end{center}
\caption{The density variable $\xcrit = \Mcrit^{2/3} \rhocrit^{1/3}$ as a function of $\beta$ for nonrotating SMSs according to Approaches I and II.  Crosses and circles denote the numerical results from Section \ref{sec:nonrotSMS:TOV}, using Approaches I and II, while the lines represent the analytical, perturbative predictions (\ref{x_nonrot_I}) and (\ref{x_nonrot_II}) from Section \ref{sec:nonrotSMS:ana} (which are identical for Approaches I and II).}
\label{fig:x_nonrot}
\end{figure}

\begin{figure}
\begin{center}
\includegraphics[width = 3in]{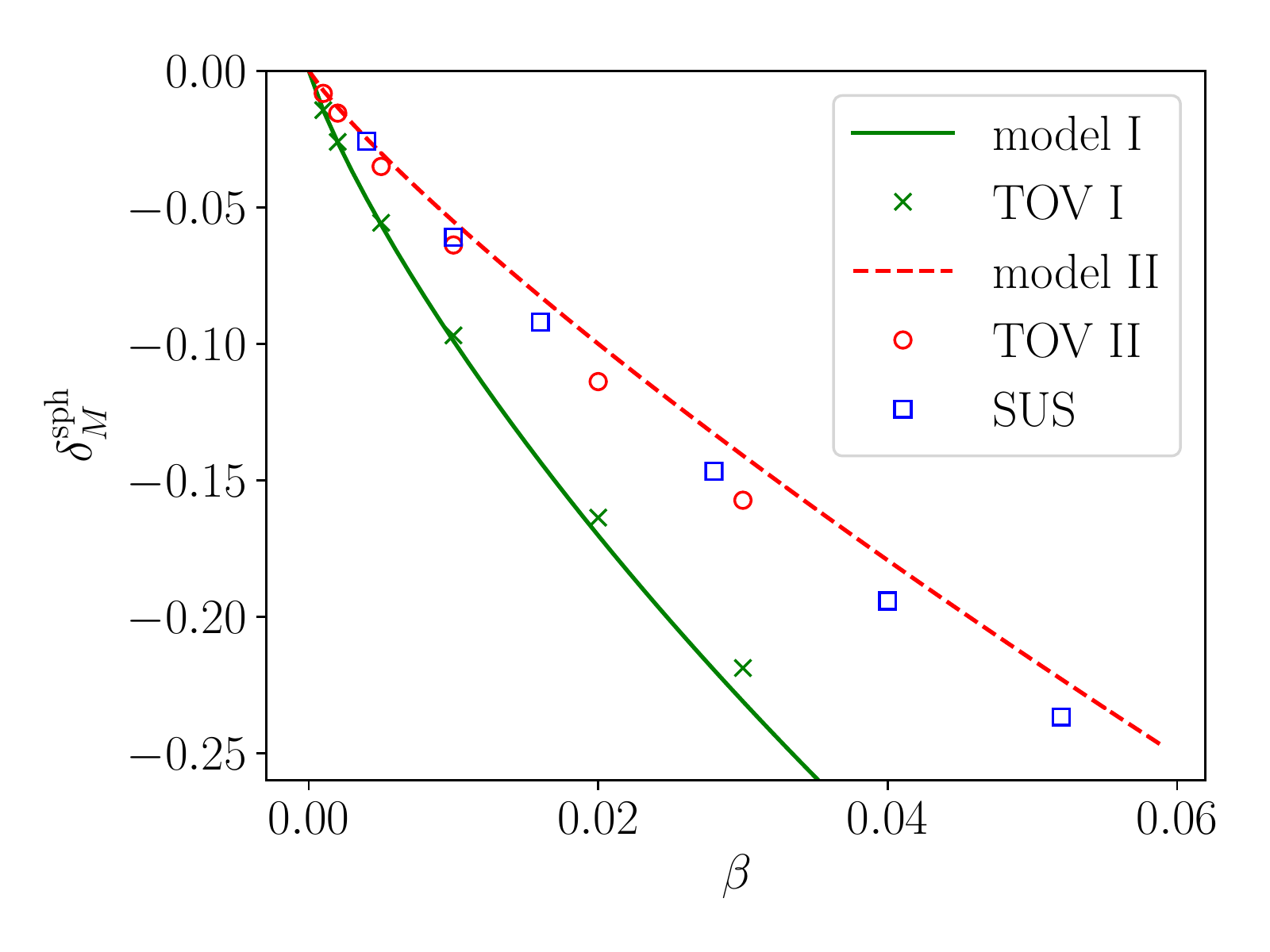}
\end{center}
\caption{The relative change in the mass $\delta_M^{\rm sph}$ (see eq.~(\ref{deltaM})) as a function of $\beta$ for nonrotating SMSs according to Approaches I and II.  Crosses and circles denote the numerical results from Section \ref{sec:nonrotSMS:TOV}, while the solid and dashed lines represent the analytical, leading-order predictions (\ref{delta_M_nonrot_I}) and (\ref{delta_M_nonrot_II}) from Section \ref{sec:nonrotSMS:ana}.  Note the non-linear behavior of the analytical predictions, which are caused by the logarithmic terms in (\ref{delta_M_nonrot_I}) and (\ref{delta_M_nonrot_II}). The squares labelled SUS represent numerical results of SUS, who adopted Approach II.}
\label{fig:M_nonrot}
\end{figure}

We construct sequences of TOV solutions and find, for given values of $\beta$, the stellar mass $\bar M$ as a function of central rest-mass density $\bar \rho_{c}$.  From these we also compute the dimensionless variable
\begin{equation} \label{x}
x \equiv \bar M^{2/3} \bar \rho_c^{1/3} = M^{2/3} \rho_c^{1/3}.
\end{equation} 
The maximum-mass configuration marks a turning point  in the curve of $\bar M$ versus $x$, at which stars become unstable to radial collapse.  We record the corresponding critical values $\bar M_{\rm crit}$ and $\xcrit$.  For $\beta = 0$, the maximum mass is found for zero density (i.e.~the curve is monotonically decreasing with $x$ and there is no turning point) and takes the well-known Newtonian value
\begin{equation} \label{M_0_sph}
\bar M_0^{\rm sph} = 4.555
\end{equation}
which we re-derive below in eq.~(\ref{M_0_sph_ks}).  Such a nonrotating SMS without gas pressure is thus unstable at all finite radii.  For small but nonzero $\beta$ we may then write
\begin{equation} \label{deltaM}
\bar M_{\rm crit} = \bar M_0^{\rm sph} ( 1 + \delta_M^{\rm sph}).
\end{equation}
In Figs.~\ref{fig:x_nonrot} and \ref{fig:M_nonrot} we show results for both $\xcrit$ and $\delta_M^{\rm sph}$ as found from Approaches I and II, which show that gas pressure can stabilize a SMS below a critical central density.  We will postpone a more detailed discussion until Section \ref{sec:nonrotSMS:comp} below.

%
\subsection{Analytical Model}
\label{sec:nonrotSMS:ana}
%

The mass and density of nonrotating SMSs can be estimated analytically from an energy variational principle \cite[see][and ST]{ZelN71}.

%
\subsubsection{General Setup}
\label{sec:nonrotSMS:ana:general}
%

\begin{table}
\begin{center}
\begin{tabular}{lll}
\hline
Coefficient & Value & Reference \\
\hline
\hline
$k_1$ & $1.7558$ &  \citet{LaiRS93} \\
$k_2$ & $0.63899$ & \citet{LaiRS93} \\
$k_3$ & $1.2041$ & \citet{LaiRS93} \\
$k_4$ & $0.918294$ & ST \\
$k_5$ & $0.331211$ & J.~C.~Lombardi Jr., 1997, priv.~comm. \\
$\ktau$ &  $- 0.45928$ & Eq.~(\ref{k_tau}) \\
$\kcdm$ & 30.0193	& Eq.~(\ref{k_CDM}) \\
$\khdm$ & $16.3262$ & Eq.~(\ref{k_HDM}); \citet{Bis98} \\
\hline
\end{tabular}
\end{center}
\caption{Values of the structure coefficients $k_i$ for $n=3$ polytropes.}
\label{table:ks}
\end{table}

We start by writing the energy as the sum of the internal energy, the gravitational potential energy, and a first post-Newtonian correction to the potential energy,
\begin{equation} \label{E_nonrot}
E = k_1 K M \rho_c^{1/n} - k_2 M^{5/3} \rho_c^{1/3} - k_4 M^{7/3} \rho_c^{2/3}
\end{equation}
\citep[see, e.g.][]{LaiRS93,BauS99b}.  Here the polytropic structure constants $k_i$ can be found from the corresponding integrals of Lane-Emden functions. We list numerical values for $n = 3$, which describes SMSs to leading order, in Table \ref{table:ks}.  It is again convenient to write the above expression in terms of dimensionless quantities.  With $\bar M = K^{-n/2} M$ and $\bar E = K^{-n/2} E$ as well as the dimensionless variable $x$ defined in (\ref{x}) we obtain
\begin{equation} \label{E_nonrot_0}
\bar E = k_1 \bar M^{1 - 2/n} x^{3/n} - k_2 \bar M x  - k_4 \bar M x^2.
\end{equation}
Equilibrium configurations can be found by setting to zero the first derivative of (\ref{E_nonrot_0}) with respect to $x$, at constant mass $\bar M$.  For a pure radiation fluid with $n = 3$ this yields the equilibrium mass
\begin{equation} \label{M_0_sph_ks}
\bar M_0^{\rm sph} = \left(\frac{k_1}{k_2} \right)^{3/2} = 4.555
\end{equation}
for $x = 0$ (cf.~(\ref{M_0_sph})).  This is the unique value of the mass in Newtonian gravitation, where the last term in (\ref{E_nonrot_0}) disappears.  Using $M = K^{3/2} \bar M$ and inserting (\ref{K}), (\ref{beta}), as well as (\ref{M_0_sph_ks}) for $\bar M$, we see that, to leading order, the mass $M$ is related to $\beta$ by
\begin{equation} \label{beta_M}
\beta \simeq 8.46 \left( \frac{M}{M_\odot} \right)^{-1/2}.
\end{equation}

The critical configuration can then be found by also setting to zero the second derivative with respect to $x$.  For a pure radiation fluid this is not possible, correctly reproducing the well-known result that all $n=3$ polytropes, in particular all stars governed by pure radiation pressure, are unstable in general relativity.   They can be stabilized by, e.g., gas pressure (e.g.~ST) and/or rotation (paper I).  We now account for the presence of gas pressure using both Approaches I and II.

%
\subsubsection{Approach I}
\label{sec:nonrotSMS:ana:appI}
%

From eq.~(\ref{eps_I}) we see that, in Approach I, the internal energy behaves like at that of an $n = 3$ polytrope plus a correction.  The total internal energy, computed from the integral
\begin{equation}
E_{\rm int} = 4 \pi \int \epsilon \, r^2 dr,
\end{equation}
therefore results in the unperturbed term $E_{\rm int,0} = k_1 \KI M \rho_c^{1/3}$ plus a correction $\Delta E_{\rm int}$.  Using (\ref{eps_I}), and following the treatment in Section 17.3 of ST, this correction can be written as
\begin{equation} \label{Delta_E}
\Delta \bar E_{\rm int} = k_1 \bar M^{1/3} x \beta \left(\frac{1}{2} \ln x + \frac{5}{6} \ln \beta + C \right). 
\end{equation}
Here we used the same dimensional rescaling as above, i.e.~$\Delta \bar E = \KI^{-n/2} \Delta E$ with $n = 3$, we have abbreviated
\begin{equation} \label{C}
C = \frac{k_\tau}{2}  - \frac{1}{3} \ln \bar M - \frac{1}{6} \ln \eta - \frac{1}{3}
\end{equation}
where $\eta \simeq 1.367 \times 10^{-4}$ is again given by (\ref{eta}), and we have defined\footnote{With this definition, the term $\tau$ defined in (ST.17.3.15) becomes $\tau = \ktau \KI \beta /2$.}
\begin{equation} \label{k_tau}
\ktau \equiv \frac{3}{k_1 | \theta'_n | \xi_n^2} \int \theta^{n+1} \ln \theta \, \xi^2 d\xi = - 0.45928. 
\end{equation}
Here $\theta$ is the Lane-Emden density function, defined so that $\rho = \rho_c \theta^n$, $\xi$ is the Lane-Emden radial function, and the last equality holds for $n=3$ polytropes.  Following the treatment in Section 17.3 in ST we assume that the mass distribution remains that of an $n = 3$ polytrope; in particular, this means that we will adopt the structure coefficients $k_i$ as listed in Table~\ref{table:ks}.

Adding (\ref{Delta_E}) to (\ref{E_nonrot_0}) and setting $n = 3$ we obtain
\begin{equation} \label{E_nonrot_I}
\bar E = k_1 \bar M^{1/3} x \left(1 + \beta \left(\frac{1}{2} \ln x + \frac{5}{6} \ln \beta + C \right) \right) 
- k_2 \bar M x - k_4 \bar M x^2.
\end{equation}
We now set the first two derivatives of (\ref{E_nonrot_I}) to zero and divide by $\bar M$, which results in the equations
\begin{align}\label{d_E_nonrot_I}
0  =  & ~k_1 \bar M^{-2/3} \left(1 + \beta \left(\frac{1}{2} \ln x + \frac{5}{6} \ln \beta + C \right) \right)  \nonumber \\
& + \frac{\beta}{2} k_1 \bar M^{-2/3} - k_2 - 2 k_4 x 
\end{align}
and 
\begin{equation} \label{dd_E_nonrot_I}
0 =  \frac{\beta}{2} k_1 \bar M^{-2/3} x^{-1} - 2 k_4.
\end{equation}
From (\ref{dd_E_nonrot_I}) we obtain
\begin{equation} \label{x_nonrot_Ia}
x_{\rm crit} = \frac{k_1}{4 k_4} \, \bar M^{-2/3} \beta.
\end{equation}
Inserting $\beta = 0$ and $x = 0$ into (\ref{d_E_nonrot_I}) we recover the result (\ref{M_0_sph_ks}), $\bar M_0^{\rm sph} = (k_1 / k_2 )^{3/2}$, which we can now insert into (\ref{x_nonrot_Ia}) to find
\begin{equation} \label{x_nonrot_I}
x_{\rm crit} = \frac{k_2}{4 k_4} \, \beta.
\end{equation}
Eq.~(\ref{x_nonrot_I}) shows that for stars of nearly the same mass, a higher ratio of gas to radiation pressure allows a higher central density, or larger compaction, before the star becomes radially unstable to collapse.  This is the stabilizing role of gas pressure.  As we will see in Section \ref{sec:nonrotSMS:ana:appII} below we will find the same result in Approach II; it is included as the solid line in Fig.~\ref{fig:x_nonrot}.  We note already the excellent agreement with the numerical results of Section (\ref{sec:nonrotSMS:TOV}) for small $\beta$.

In order to obtain an expression for the correction to the mass at the critical point we insert (\ref{dd_E_nonrot_I}), (\ref{x_nonrot_I}) and
\begin{equation} \label{M_sph_expansion}
\bar M = \bar M_0^{\rm sph} (1 + \delta_M^{\rm sph})
\end{equation}
into (\ref{d_E_nonrot_I}) and expand to leading order to obtain
\begin{equation} \label{delta_M_nonrot_I}
\delta_M^{\rm sph,I} = \left( \frac{3}{4} \ln \frac{k_2}{4 k_4} + 2 \ln \beta + \frac{3}{2} C \right) \beta,
\end{equation}
where the superscript `I' refers to Approach I.  Since $\delta_M^{\rm sph}$ already describes a leading-order correction, we may replace $\bar M$ with $\bar M_0^{\rm sph}$ in $C$.  The prediction (\ref{delta_M_nonrot_I}) is included in Fig.~\ref{fig:M_nonrot} as a solid line.  Note that, for small $\beta$, this expression is dominated by the term proportional to $\beta \ln \beta$.  

%
\subsubsection{Approach II}
\label{sec:nonrotSMS:ana:appII}
%

In Approach II we simply adopt $n = n_1$, as given in (\ref{n_1}), in the energy functional (\ref{E_nonrot_0}), which then becomes
\begin{equation} \label{E_nonrot_II}
\bar E = k_1 \bar M^{1/3 - \beta/3} x^{1 + \beta/2} - k_2 \bar M x - k_4 \bar M x^2.
\end{equation}
We now use the dimensional rescaling introduced above with $\KII$ and $n_1$, e.g.~$\bar E = \KII^{-n_1/3} E$, where we approximate $\KII$ as a constant (see eq.~(\ref{K_II})).  Setting the first two derivatives of (\ref{E_nonrot_II}) to zero and dividing by $\bar M$ now yields 
\begin{equation} \label{d_E_nonrot_II}
0  =  \left(1 + \frac{\beta}{2} \right) k_1 \bar M^{- 2/3 - \beta/3} x^{\beta/2} - k_2 
- 2 k_4  x 
\end{equation}
and
\begin{equation} \label{dd_E_nonrot_II}
0  =  \frac{\beta}{2} \left(1 + \frac{\beta}{2} \right) k_1 \bar M^{- 2/3 - \beta/3} x^{\beta/2 - 1} 
- 2 k_4.  
\end{equation}
As in Approach I we will ignore changes in the structure constants $k_i$.

Combining eqs.~(\ref{d_E_nonrot_II}) and (\ref{dd_E_nonrot_II}) we obtain
\begin{equation} 
\frac{\beta}{2} k_2 - 2 \left(1 - \frac{\beta}{2} \right) k_4 x = 0,
\end{equation}
which, to leading order, yields
\begin{equation} \label{x_nonrot_II}
x_{\rm crit} = \frac{k_2}{4 k_4} \beta,
\end{equation}
the exact same result (\ref{x_nonrot_I}) that we previously obtained in approach~I.

We can now insert (\ref{x_nonrot_II}) into (\ref{d_E_nonrot_II}) to obtain an expression for the change in the mass.  We again use the expansion (\ref{M_sph_expansion}).  Expanding the exponents in (\ref{d_E_nonrot_II}) to linear order we find
\begin{align} \label{M_expand_nonrot}
\bar M^{-2/3 - \beta/3} & \simeq (\bar M_0^{\rm sph})^{-2/3 - \beta/3} \left(1 - \frac{2}{3} \delta_M^{\rm sph} \right) \nonumber \\
& = (\bar M_0^{\rm sph})^{-2/3} \exp\left( - \frac{\beta}{3} \ln \bar M_0^{\rm sph} \right) \left(1 - \frac{2}{3} \delta_M^{\rm sph} \right)  \nonumber \\
& \simeq (\bar M_0^{\rm sph})^{-2/3} \left(1 - \frac{\beta}{3} \ln \bar M_0^{\rm sph} - \frac{2}{3} \delta_M^{\rm sph}\right)
\end{align}
as well as
\begin{align} \label{x_expand_nonrot}
x^{\beta/2} & =  \left(\frac{k_2}{4 k_4} \beta \right)^{\beta/2}  
=  \exp\left( \frac{\beta}{2} \ln \left( \frac{k_2}{4 k_4} \right) \right) \exp\left( \frac{\beta}{2} \ln \beta \right) \nonumber \\
& \simeq 1 + \frac{\beta}{2} \ln  \left( \frac{k_2}{4 k_4} \right)  + \frac{\beta}{2} \ln \beta.
\end{align}
Inserting these expansions into (\ref{d_E_nonrot_II}) then yields
\begin{align}
0 = k_1 (\bar M_0^{\rm sph})^{-2/3} \Bigg( & 1 + \frac{\beta}{2} - \frac{\beta}{3} \ln \bar M_0^{\rm sph} - \frac{2}{3} \delta_M  \nonumber \\
& + \frac{\beta}{2} \ln \frac{k_2}{4 k_4} + \frac{\beta}{2} \ln \beta \Bigg) - k_2 - k_2 \frac{\beta}{2},
\end{align}
Not surprisingly, this equation yields the Newtonian mass (\ref{M_0_sph_ks}) to zeroth order.  The next leading order terms can be solved for $\delta_M^{\rm sph}$ to yield
\begin{equation} \label{delta_M_nonrot_II}
\delta_M^{\rm sph,II} = \left( \frac{3}{4} \ln \frac{k_2}{4 k_4}  + \frac{3}{4} \ln \beta - \frac{1}{2} \ln \bar M_0^{\rm sph}  \right) \beta,
\end{equation}
where we have used (\ref{M_0_sph_ks}) to eliminate $k_2$.  This result is included in Fig.~\ref{fig:M_nonrot} as a dashed line.   Note that this result is different from that obtained in Approach I, eq.~(\ref{delta_M_nonrot_I}), as we will discuss in more detail in the following Section.

%
\subsection{Comparisons}
\label{sec:nonrotSMS:comp}
%

Figs.~\ref{fig:x_nonrot} and \ref{fig:M_nonrot} show the numerical results of Section \ref{sec:nonrotSMS:TOV} and the analytical results of Section \ref{sec:nonrotSMS:ana} for both Approaches I and II.  We emphasize again that all calculations employed here, both numerical and analytical, treat the effects of gas pressure only approximately.  Approach I is based on a Taylor-expansion that approximates the total entropy $s$ to be constant to leading-order only (see Section \ref{sec:thermo:appI}), while Approach II approximates the EOS as polytropic, leading to a polytropic constant that is only approximately constant (see Section \ref{sec:thermo:appII}).  We believe that the nature of the approximation is more fundamental in Approach I. However, as we will discuss in more detail below, we find that both approaches make identical predictions for some key dimensionless quantities at the critical point.

We have already noted that Approaches I and II give identical results (\ref{x_nonrot_I}) and (\ref{x_nonrot_II}) for the dimensionless density variable $x_{\rm crit}$; the two lines representing these results therefore lie on top of each other in Fig.~\ref{fig:x_nonrot}.  In this figure we see that this analytical prediction also agrees very well with the numerical results for both Approaches I and II, at least to leading order in $\beta$.  Since all calculations included in the figure are accurate to leading order only, the apparently better agreement between the analytical prediction and the numerical results for Approach II for larger $\beta$ is presumably coincidental.  

We already noted that Approaches I and II make different predictions for the change in the rescaled mass $\delta_M^{\rm sph}$; see eqs.~(\ref{delta_M_nonrot_I}) and (\ref{delta_M_nonrot_II}).  In Fig.~\ref{fig:M_nonrot} we include the two predictions with a solid and dashed lines and find that, for small $\beta$, they each agree very well with the respective numerical results for Approach I and II.  In fact, it is not surprising that the two different approaches result in different values of $\bar M$, since the two different approaches adopt two different rescalings between $\bar M$ and the physical mass $M$:  In Approach I we have $M = \KI^{3/2} \bar M$, and in Approach II $M = \KII^{n_1/2} \bar M$.  Even inserting the different constants $\KI$ (\ref{K_I}) and $\KII$ (\ref{K_II}) results in some ambiguities, since $\KI$ assumes that the total entropy $s$ is constant, while $\KII$ depends on the radiation entropy $s_r$ and, weakly, on the density $\rho_0$.  In order to avoid these ambiguities we will therefore focus on dimensionless quantities that do not depend on $K$.  We have already seen that both approaches make identical predictions for $x_{\rm crit}$, which relates the central density $\rho_0$ to the stellar mass $M$.  We will similarly see in Section \ref{sec:gas} that, for rotating stars, both approaches also make identical predictions for the change in the dimensionless parameters that we are primarily interested in, namely the angular momentum $j_{\rm crit} = (J/M^2)_{\rm crit}$, and the compaction $(R/M)_{\rm crit}$, both evaluated at the maximally rotating critical configuration.

For completeness, though, we also include in Fig.~\ref{fig:M_nonrot} the numerical results of SUS, who adopted Approach II to model the effects of gas pressure.   Since SUS do not provide values for $\beta = 0$, i.e.~$\Gamma = 4/3$, we extrapolate the numerical values $\bar M^{\rm sph}$ listed in their Table I to $\beta = 0$, which yields their value of the Newtonian, spherical mass $\bar M_0^{\rm sph}$.  We then compute $\delta_M^{\rm sph} = (\bar M^{\rm sph} - \bar M^{\rm sph}_0) / \bar M^{\rm sph}_0$.  As expected, their values agree well with our numerical values for Approach II, and, for small $\beta$, both agree well with our analytical prediction for Approach II.

Finally we demonstrate that our results reproduce those of ST, who adopt Approach I to compute the density of nonrotating SMSs in their Section 17.4, but use a different notation.  Inserting the definition (\ref{x}) into (\ref{x_nonrot_I}) we obtain
\begin{equation}
\rho_{\rm crit} = \left( \frac{1}{4} \frac{k_2}{k_4} \right)^3 \frac{\beta^3}{M^2}
 = \left( \frac{8.5}{4} \frac{k_2}{k_4} \frac{c^2}{G} \right)^3 \left( \frac{M_\odot}{M} \right)^{7/2} \frac{1}{M_\odot^2}, 
\end{equation}
where have used the relation (\ref{beta_M}) in the last step, and where we have also inserted appropriate powers of $c$ and $G$ to obtain an expression in cgs units.  Inserting values for the latter, as well as the solar mass $M_\odot$, we obtain 
\begin{equation} \label{rho_crit_phys}
\rho_{\rm crit} = 1.92 \times 10^{-3} 
  \left( \frac{10^6 M_\odot}{M} \right)^{7/2} \frac{\mbox{g}}{\mbox{cm}^3}, 
\end{equation}
very similar to the value provided in eq.~(ST.17.4.7).   Using $M = 4\pi \rho_{\rm ave} R^3/3$ as well as $\rho_c / \rho_{\rm ave} \simeq 54.18$ for an $n=3$ polytrope we also have
\begin{equation} \label{RoverM_crit_phys}
\left( \frac{R}{M} \right)_{\rm crit} = 1.59 \times 10^3 \left( \frac{M}{10^6 M_{\odot}} \right)^{1/2}
\end{equation}
(compare (ST.17.4.11)).  Eq.~(\ref{rho_crit_phys}) gives the critical density for the onset of radial collapse of a spherical SMS with both radiation and gas pressure and (\ref{RoverM_crit_phys}) its value of $R/M$: stars with $\rho < \rho_{\rm crit}$ and $R/M > (R/M)_{\rm crit}$ are stable, while those with $\rho > \rho_{\rm crit}$ and $R/M < (R/M)_{\rm crit}$ are unstable to collapse.

%
\section{Perturbative effects on the stability of rotating SMSs}
\label{sec:rotSMS}
%

In this Section we extend the analytical model of Section \ref{sec:nonrotSMS:ana:general} to develop a general framework for treating perturbative effects on uniformly rotating SMSs.  We first review the model and results obtained in Paper I for the critical configuration in the absence of other perturbations in Sections \ref{sec:rotSMS:review} and \ref{sec:rotSMS:background}, and then introduce general expressions for perturbations of this critical solution in Section \ref{sec:rotSMS:perturb}.  We will leave the nature of the perturbation unspecified in this Section.  In Section \ref{sec:gas} we will then apply this formalism to gas pressure and in Section \ref{sec:others} to other effects, all in the presence of rotation.

%
\subsection{Review of the analytical model}
\label{sec:rotSMS:review}
%

In order to model rotating SMSs we include in the energy functional (\ref{E_nonrot}) both a rotational energy term as well as a second post-Newtonian correction to the Newtonian potential energy,
\begin{align} \label{energy}
E = & ~ k_1 K M \rho_c^{1/n} - k_2 M^{5/3} \rho_c^{1/3} + k_3 j^2 M^{7/3} \rho_c^{2/3} \nonumber \\
& - k_4 M^{7/3} \rho_c^{2/3} - k_5 M^3 \rho_c
\end{align}
(see e.q.~(20) in paper I, hereafter (I.20)).  Here $j = J/M^2$ is a dimensionless measure of the angular momentum $J$, and the structure constants $k_i$ are again listed in Table \ref{table:ks} for an $n = 3$ polytrope.  As in Section \ref{sec:nonrotSMS:ana} we will assume that these structure constants remain unchanged in all our perturbative calculations.  Even though the second post-Newtonian correction, the last term in the above expression, takes a very small value, its inclusion is crucial for determining the critical configuration, i.e.~the onset of instability \citep[see][Paper I]{ZelN71}.  Using the same dimensional rescaling as before, e.g.~$\bar M \equiv K^{-n/2} M$ and $\bar E \equiv K^{-n/2} E$, as well as the definition (\ref{x}),
we can rewrite (\ref{energy}) as
\begin{align} \label{energy1}
\bar E = & ~ k_1 \bar M^{1 - 2/n} x^{3/n} - k_2 \bar M x  + k_3 j^2 \bar M x^2 \nonumber \\
&  - k_4 \bar M x^2 - k_5 \bar M x^3.
\end{align}
We note that in Paper I we used the alternative scaling $\tilde x \equiv \bar \rho_c^{1/3} = \bar M^{-2/3} x$ for $n=3$.  The advantage of the scaling (\ref{x}), $x = \bar M^{2/3} \bar \rho_c^{1/3}$,  is that the dimensionless mass $\bar M$ now appears with the same power in all terms except in the first.

Equilibrium configurations can be found by setting to zero the derivative of the energy (\ref{energy1}) with respect to $x$ (at constant mass and angular momentum), 
\begin{align} \label{energy_deriv_1}
0 = \frac{\partial \bar E}{\partial x} = & ~ (3/n) k_1 \bar M^{1 - 2/n} x^{3/n - 1} - k_2 \bar M  + 2 k_3 j^2 \bar M x \nonumber \\
& - 2 k_4 \bar M x - 3 k_5 \bar M x^2.
\end{align}
The onset of instability occurs at turning points of the equilibrium sequences, i.e.~at points at which the second derivative with respect to $x$ vanishes,
\begin{align} \label{energy_deriv_2}
0 = \frac{\partial^2 \bar E}{\partial x^2} = &  ~(3/n)(3/n - 1) k_1 \bar M^{1 - 2/n} x^{3/n - 2} + 2 k_3 j^2 \bar M \nonumber \\
& - 2 k_4 \bar M - 6 k_5 \bar M x.
\end{align}
For $n=3$, eqs.~(\ref{energy_deriv_1}) and (\ref{energy_deriv_2}) are equivalent to eqs.~(I.21) and (I.22).

The ratio between the rotational kinetic and Newtonian potential energies can be computed from the second and third terms in eq.~(\ref{energy}),
\begin{equation} \label{ToverW}
\frac{T}{|W|} = \frac{k_3 j^2 x}{k_2}
\end{equation}
(compare eq.~(I.27)).  Adopting the Roche model, we can compute the rotational kinetic energy at mass shedding from $T = (1/2) I \Omega^2_{\rm shed}$.   Consistent with our assumption that the structure constants $k_i$ remain unchanged,  we will assume that the moment of inertia $I$ always remains that of the nonrotating $n=3$ polytrope, $I \simeq (2/3) \, 0.1130 M R_p^2$ (see (I.18)), and where $\Omega_{\rm shed} = (2/3)^{3/2} (M/R_p^3)^{1/2}$ (see (I.17)).  We similarly assume that the potential energy remains that of the nonrotating $n=3$ polytrope, $|W| = (3/2) M^2/R_p$.  Combining these terms we find
\begin{equation} \label{roche}
\left( \frac{T}{|W|} \right)_{\rm shed} = 7.44 \times 10^{-3}
\end{equation}
(see eq.~(I.19)), which can be inserted into (\ref{ToverW}) for a star uniformly rotating at mass-shedding, the maximum spin rate.   Note that the small value of $T/|W|$ at mass-shedding for an $n=3$ polytope (due to its large central mass concentration) justifies ignoring changes in the shape of the bulk of the mass in computing the energy functional, and treating the star as spherical to lowest order.

Noting that we may also write the kinetic rotational energy as $T = (1/2) J^2/I$ we find
\begin{equation}
\frac{T}{|W|} = 4.425 \frac{M}{R_p} j^2,
\end{equation}
which can then be inverted to yield an expression for the compaction $R_p/M$,
\begin{equation} \label{RoverM}
\frac{R_p}{M} = 4.425 \frac{j^2}{T/|W|}.
\end{equation}

Eqs.~(\ref{energy_deriv_1}), (\ref{energy_deriv_2}) and (\ref{ToverW}), with $T/|W|$ given by (\ref{roche}), now provide three equations for the three unknown parameters $x$, $\bar M$ and $j$ describing the critical configuration of marginally stable SMSs rotating at mass-shedding.  In the following Sections we will solve these equations perturbatively, adopting as the unperturbed background solution the critical configuration of an $n=3$ polytrope as derived in Paper I.

%
\subsection{The unperturbed critical configuration}
\label{sec:rotSMS:background}
%

In Paper I we solved eqs.~(\ref{energy_deriv_1}), (\ref{energy_deriv_2}) and (\ref{ToverW}) under the idealized assumption that the SMS is dominated by a radiation fluid, so that $n = 3$.  In that case eq.~(\ref{energy_deriv_2}) yields
\begin{equation} \label{x0}
x_0 = \frac{k_3 j_0^2 - k_4}{3 k_5}
\end{equation}
(compare (I.23)), where we have introduced the subscript `0' to denote parameters describing the critical configuration of the unperturbed $n=3$ polytrope.  As an immediate consequence, we see that rotation alone can stabilize the star only for angular momenta $j_0$ greater than a minimum angular momentum
\begin{equation} \label{j_min}
j_{\rm min} = \left( \frac{k_4}{k_3} \right)^{1/2} = 0.8733.
\end{equation}
Using this result we can rewrite (\ref{x0}) as 
\begin{equation} \label{x2}
x_0 = \frac{k_3}{3 k_5}(j_0^2 - j_{\rm min}^2)
\end{equation}
and eq.~(\ref{energy_deriv_1}) as
\begin{equation} \label{M0_1}
k_1 \bar M_0^{-2/3} = k_2 - 2 k_3 (j_0^2 - j_{\rm min}^2) x_0 + 3 k_5 x_0^2.
\end{equation}
Inserting (\ref{x2}) into the latter then yields
\begin{equation} \label{M0_2}
\bar M_0^{2/3} = k_1 \left( k_2 - \frac{k_3^2}{3 k_5} \left(j_0^2 - j_{\rm min}^2 \right)^2 \right)^{-1}
\end{equation}
(see eq.~(I.25)).  Inserting eq.~(\ref{x2}) into (\ref{ToverW}) results in a quadratic equation for $j_0^2$,
\begin{equation} \label{j_0}
j_0^4 - j_{\rm min}^2 j_0^2 - \frac{3 k_2 k_5}{k_3^2} \, \frac{T}{|W|} = 0,
\end{equation}
which can be solved for $j_0$ as a function of $T/|W|$.  This solution provides the dimensionless angular momentum of the critical configuration,
\begin{equation} \label{JM20}
\left( \frac{J}{M^2} \right)_{\rm crit,0} = j_0 = 0.876.
\end{equation}
Inserting $j_0$ together with (\ref{roche}) into (\ref{RoverM}) then yields 
\begin{equation} \label{RoverM0}
\left( \frac{R_p}{M} \right)_{\rm crit,0}  \simeq 456
\end{equation}
for the analytical model.  In the Roche model the equatorial radius $R_{\rm eq}$ is related to the polar radius $R_p$ by $R_{\rm eq} = 3 R_p / 2$, hence
\begin{equation} 
\left( \frac{R_{\rm eq}}{M} \right)_{\rm crit,0}  \simeq 684.
\end{equation}
In addition to this analytical model calculation we also performed a fully relativistic, numerical calculation in Paper I, which resulted in $j_0 \simeq 0.97$ and $R_p/M \simeq 427$.  From eq.~(\ref{x2}) for $x_0$ we also find the critical central density
\begin{align} \label{rho_crit_0}
\rho_{\rm crit,0} &=  \left( \frac{c^2}{G} \right)^3 \frac{x_0^3}{M_\odot^2}  \left( \frac{M_{\odot}}{M} \right)^2 
= 8.7 \times 10^{-2} \left( \frac{10^6 M_{\odot}}{M} \right)^2 \frac{\mbox{g}}{\mbox{cm}^3}.
\end{align}
Eqs.~(\ref{RoverM0}) and (\ref{rho_crit_0}) should be compared with eqs.~(\ref{RoverM_crit_phys}) and (\ref{rho_crit_phys}).  The former two describe the physical properties of the critical configuration of a SMS star stabilized by uniform rotation, while the latter two describe those of the critical configuration of a SMS stabilized by gas pressure.

\begin{table*}
\begin{center}
\begin{tabular}{llllll}
\hline
\hline
Calculation	& $(R_p/M)_{\rm crit}$ &  $x_0$ & $j_{\rm min}$		& $j_0$	& $\bar M_0$ \\ 
\hline
Paper I; analytical	& 456 & $5.15 \times 10^{-3}$ & 0.8733	&	0.8757	& 4.5551 \\
Paper I; numerical	& 427 & $5.26 \times 10^{-3}$ & 0.882	&	0.97		& 4.57 \\
SUS	& & & 		&	0.921	& 4.56 \\
\hline
\end{tabular}
\end{center}
\caption{Characteristic parameters of the unperturbed critical configuration}
\label{table:crit}
\end{table*}

In Table \ref{table:crit} we list some of the characteristic parameters of the unperturbed critical configuration, as found in Paper I from both the analytical model calculation and the numerical simulations (cf.~Table 2 in Paper I).    We also include numerical values for $j_0$ and $\bar M_0$ as computed from the values provided by SUS.  As discussed in Section \ref{sec:nonrotSMS:comp}, we compute these values by extrapolating the data listed in their Table 1 to $n = 3$.  

Eqs.~(\ref{JM20}) and (\ref{RoverM0}) provide the unique parameters of the critical configuration of a uniformly rotating SMS, supported by pure radiation pressure and spinning at the mass-shedding limit.  The subsequent gravitational collapse to a black hole is therefore also unique, up to the overall scaling with mass, and the emitted gravitational wave-signal may hence serve as a ``standard-siren" for future space-based gravitational wave detectors.  In the following Sections we will consider perturbations of this critical configurations in order to explore the regimes in which this universality is affected by the presence of gas pressure, magnetic fields, dark matter and dark energy.

%
\subsection{Perturbative treatment: general setup}
\label{sec:rotSMS:perturb}
%

We will account for perturbations of the idealized assumptions of Section \ref{sec:rotSMS:background} (and Paper I) by including new terms in the energy functional (\ref{energy}).   These terms therefore lead to perturbations in the variables that characterize the critical configuration, namely the density variable $x$, the angular momentum $j$, and mass $\bar M$, given by solutions to eqs.~(\ref{energy_deriv_1}), (\ref{energy_deriv_2}) and (\ref{ToverW}).  In our perturbative approach we write these deviations as
\begin{align} 
x & =  x_0 ( 1 + \delta_x), \label{delta_x} \\
j & =  j_0 ( 1 + \delta_j), \label{delta_j} \\
\bar M  & = \bar M_0 ( 1 + \delta_M), \label{delta_M} 
\end{align}
where $x_0$, $j_0$ and $\bar M_0$ are the unperturbed parameters of Section \ref{sec:rotSMS:background}.  We are assuming that the onset of instability is still dominated by the interplay between radiation pressure, rotation and post-Newtonian corrections to the potential energy, as outlined in Section \ref{sec:rotSMS:review}, and that all corrections (e.g.~gas pressure in the limit $\beta \ll 1$) affect this onset of instability only perturbatively.  Other approaches are possible, of course, but are not what we pursue in this paper.  

Specific expressions for $\delta_x$, $\delta_j$ and $\delta_M$ will depend on the specific effects considered, and will be derived in the following Sections.  Eq.~(\ref{ToverW}), however, is independent of the energy function itself, and, thanks to our rescaling (\ref{x}), also does not depend on $\bar M$.  Assuming that $T/|W|$ remains given by (\ref{roche}), which is consistent with our assumption that the structure constants $k_i$ remain unchanged, we may therefore insert both (\ref{delta_x}) and (\ref{delta_j}) into (\ref{ToverW}) and expand to linear order to obtain 
\begin{equation} \label{ToverW2}
\frac{T}{|W|} = \frac{k_3}{k_2} j_0^2 x_0 (1 + 2 \delta_j + \delta_x).
\end{equation}
Evidently, to linear order we must always have
\begin{equation} \label{delta_j_2}
\delta_x = - 2 \, \delta_j.
\end{equation}
We then insert (\ref{delta_j}) into (\ref{RoverM}) to obtain
\begin{equation} \label{RoverM1}
\left( \frac{R_p}{M} \right)_{\rm crit} = 
\left( \frac{R_p}{M} \right)_{\rm crit,0} ( 1 + 2 \delta_j).
\end{equation}
to linear order.   From (\ref{delta_j}) we also have 
\begin{equation} \label{JoverM21}
\left( \frac{J}{M^2} \right)_{\rm crit} = 
\left( \frac{J}{M^2} \right)_{\rm crit,0} ( 1 + \delta_j).
\end{equation}
The above expressions can now be used to determine the change in the angular momentum $j$ and the critical compaction $R_p/M$ once $\delta_x$ has been found.

%
\section{Effects of gas pressure}
\label{sec:gas}
%

In this Section we return to the effects of gas pressure, and consider its effect on the stability of maximally rotating SMSs using both Approaches I and II (Sections \ref{sec:gas:appI} and \ref{sec:gas:appII}).  We compare with the numerical results of SUS in Section \ref{sec:gas:comp}, and evaluate our results to compute changes in the physical parameters of critically spinning SMSs in Section \ref{sec:gas:results}.

%
\subsection{Approach I}
\label{sec:gas:appI}
%

Following Section \ref{sec:nonrotSMS:ana:appI} we account for gas pressure by adopting (\ref{energy1}) with $n = 3$, as well as $K = \KI$ in the dimensional rescaling, but adding the correction (\ref{Delta_E}) to the internal energy.  The energy functional then becomes
\begin{align} \label{energy_appI}
\bar E  = &~ k_1 \bar M^{1/3} x + k_1 \bar M^{1/3} x \beta \left( \frac{1}{2} \ln x + \frac{5}{6} \ln \beta + C \right)  \nonumber \\
& - k_2 \bar M x  + k_3 j^2 \bar M x^2 - k_4 \bar M x^2 - k_5 \bar M x^3,
\end{align}
where $C$ is given by (\ref{C}).  Setting the first two derivatives to zero we obtain
\begin{align} \label{denergy_appI}
0 = &  ~k_1 \bar M^{2/3}  k_1 \bar M^{-2/3} \beta \left( \frac{1}{2} \ln x + \frac{5}{6} \ln \beta + C \right) 
\nonumber \\
& + \frac{1}{2} k_1 \bar M^{-2/3}  - k_2  + 2 k_3 j^2  x - 2 k_4 x - 3 k_5  x^2
\end{align}
and 
\begin{equation} \label{ddenergy_appI}
0  = \frac{1}{2} k_1 \bar M^{-2/3} \beta x^{-1} + 2 k_3 j^2 - 2 k_4 - 6 k_5 x.
\end{equation}
We now insert (\ref{delta_x}) and (\ref{delta_j}) into (\ref{ddenergy_appI}) to find, to leading order,
\begin{equation} \label{ddenergy_appI_pert}
0  = \frac{1}{2}  \bar M_0^{-2/3} x_0^{-1} \beta + 2 k_3 j_0^2 (1 + 2 \delta_j) - 2 k_3 j_{\rm min}^2 - 6 k_5 x_0 ( 1 + \delta_x),
\end{equation}
where we have used (\ref{j_min}).   Not surprisingly, the zeroth-order term gives us (\ref{x2}) again.  Using (\ref{delta_j_2}), the linear terms can be solved to yield an expression for the change in $j$,
\begin{equation} \label{delta_j_appI}
\delta_j = - \frac{k_1}{8 k_3} \, \frac{1}{\bar M_0^{2/3} (2 j_0^2 - j_{\rm min}^2) x_0} \, \beta.
\end{equation}
This expression can be inserted into (\ref{RoverM1}) and (\ref{JoverM21}) to find a change in the dimensional parameters $R_p/M$ and $J/M^2$ -- the quantities that we are primarily interested in.  We will postpone a discussion of these results, though, until Section \ref{sec:gas:comp}.  The change in the density variable $x$ can also be found from (\ref{delta_j_appI}) with the help of (\ref{delta_j_2}).  

Keeping in mind the discussion of $\delta_M$ in Section \ref{sec:nonrotSMS:comp}, we can compute this quantity as follows.  We first insert (\ref{ddenergy_appI}) into (\ref{denergy_appI}) to obtain
\begin{equation} \label{appI_1}
0 = k_1 \bar M^{-2/3} + k_1 \bar M^{-2/3} \beta \left( \frac{1}{2} \ln x + \frac{5}{6} \ln \beta + C \right) 
 - k_2 + 3 k_5 x^2.
\end{equation}
We then insert (\ref{delta_x}) and (\ref{delta_M}) and expand to linear order, which yields
\begin{equation}
\delta_M^I = \left( \frac{3}{4} \ln x_0 + \frac{5}{4} \ln \beta + \frac{3}{2} C \right) \beta + \frac{9 k_5 \bar M_0^{2/3} x_0^2}{k_1} \delta_x,
\end{equation}
or, with (\ref{delta_j_2}) and (\ref{delta_j_appI}),
\begin{equation} \label{delta_M_appI}
\delta_M^I = \left( \frac{3}{4} \ln x_0 + \frac{5}{4} \ln \beta + \frac{3}{2} C  + \frac{9 k_5}{4 k_3} \, \frac{x_0}{2 j_0^2 - j_{\rm min}^2}  \right) \beta.
\end{equation}
Here we have introduced a superscript ``I" in order to distinguish this result from the corresponding result (\ref{delta_M_appII}) in Approach II.  We will again postpone a discussion until after we have presented results for Approach II in the next Sections.

%
\subsection{Approach II}
\label{sec:gas:appII}
%

As in Section \ref{sec:nonrotSMS:ana:appII} we now account for gas pressure by adopting the approximate polytropic index $n = n_1$ as given by (\ref{n_1}), using $K = \KII$ (approximated as a constant) in the dimensional rescaling.  The first two derivatives (\ref{energy_deriv_1}) and (\ref{energy_deriv_2}) of the energy function (\ref{energy1}) can then be written as 
\begin{equation} \label{energy_deriv_1_pres}
0  = ~ \left(1 + \frac{\beta}{2} \right) k_1 \bar M^{- 2/3 - \beta/3} x^{\beta/2} - k_2 
 + 2 k_3 (j^2  - j_{\rm min}^2 ) x  - 3 k_5  x^2
\end{equation}
and
\begin{equation} \label{energy_deriv_2_pres}
0 = \left(1 + \frac{\beta}{2} \right) \frac{\beta}{2} k_1 \bar M^{- 2/3 - \beta/3} x^{\beta/2 - 1} 
  + 2 k_3 (j^2 - j_{\rm min}^2) - 6 k_5  x,
\end{equation}
where we have used (\ref{j_min}).  These two equations can then be combined to eliminate the first terms, leading to
\begin{equation} \label{energy_deriv_combined_pres}
\frac{\beta}{2} k_2 
+ 2 \left(1 - \frac{\beta}{2} \right) k_3 (j^2 - j_{\rm min}^2)  x -
6 \left( 1 - \frac{\beta}{4} \right) k_5  x^2  = 0.
\end{equation}
Inserting (\ref{delta_x}) and (\ref{delta_j}), the leading-order correction terms become
\begin{align} 
0 = & ~\frac{\beta}{2} k_2  + 2 k_3  j_0^2 x_0  \left( 2 \delta_j + \delta_x - \frac{\beta}{2} \right) \\
& - 2 k_3 j_{\rm min}^2 x_0 \left( \delta_x - \frac{\beta}{2} \right) - 6 k_5 x_0^2 \left(2 \delta_x - \frac{\beta}{4} \right). \nonumber 
\end{align}
We now use (\ref{delta_j_2}) to eliminate $\delta_x$ and obtain
\begin{align}
& \left(4 k_3 j_{\rm min}^2 x_0  + 24 k_5 x_0^2 \right) \, \delta_j =  \\
  & ~~~~~~~~~~~ \left( - \frac{1}{2} k_2 + k_3 (j_0^2 - j_{\rm min}^2) x_0 - \frac{3}{2} k_5 x_0^2 \right)  \,\beta. \nonumber
\end{align}
Simplifying the left-hand side with (\ref{x2}), and the right-hand side with (\ref{M0_1}) yields the result
\begin{equation} \label{delta_j_appII}
\delta_j = - \frac{k_1}{8 k_3} \, \frac{1}{ \bar M_0^{2/3} (2 j_0^2 - j^2_{\rm min}) x_0} \, \beta,
\end{equation}
which is identical to the result (\ref{delta_j_appI}) that we found for Approach I.  In particular, this means that both approaches predict identical changes in the parameters $R_p/M$ and $J/M^2$ of the critical configuration, see eqs.~(\ref{RoverM1}) and (\ref{JoverM21}), as well as for $\delta_x$, see eq.~(\ref{delta_j_2}).

In order to obtain an expression for the correction $\delta_M$ of the rescaled mass $\bar M$ we expand eq.~(\ref{energy_deriv_1_pres}).  For the expansion of the mass term $\bar M^{-2/3 - \beta/3}$ we can use the expansion (\ref{M_expand_nonrot}), but, unlike in (\ref{x_expand_nonrot}), we now expand $x$ about $x_0$ rather than zero, so we now have
\begin{equation} \label{x_expand}
x^{\beta/2} = e^{(\beta/2) \ln x} = 1 + \frac{\beta}{2} \ln x_0 
\end{equation}
to linear order.  Inserting these, together with (\ref{delta_x}) and (\ref{delta_j}), into (\ref{energy_deriv_1_pres}) then yields
\begin{align} 
0 = & ~ k_1 \bar M_0^{-2/3} \left(1 + \frac{\beta}{2} - \frac{\beta}{3} \ln \bar M_0  + \frac{\beta}{2} \ln x_0  - \frac{2}{3} \delta_M \right) \nonumber \\
& - k_2 + 2 k_3 j_0^2 x_0 \left(1 + 2 \delta_j + \delta_x \right) \nonumber \\
& - 2 k_3 j_{\rm min}^2 x_0 \left(1 + \delta_x \right) - 3 k_5 x_0^2 \left(1 + 2 \delta_x \right).
\end{align}
Not surprisingly, the zeroth-order terms reproduce eq.~(\ref{M0_1}).  The leading-order correction terms can be simplified by using (\ref{delta_j_2}) to write $\delta_x$ in terms of $\delta_j$, and (\ref{delta_j_appII}) to write $\delta_j$ in terms of $\beta$.  Also using (\ref{x2}) we then find
\begin{equation} \label{delta_M_appII}
\delta_M^{II} = \left( \frac{3}{4} - \frac{1}{2} \ln \bar M_0 + \frac{3}{4} \ln x_0 - \frac{3}{4} \frac{j_0^2}{2 j_0^2 - j_{\rm min}^2} \right) \beta.
\end{equation}
Given our discussion in Section \ref{sec:nonrotSMS:comp} it is not surprising that this result is not identical with that found for Approach I, eq.~(\ref{delta_M_appI}).

%
\subsection{Comparisons}
\label{sec:gas:comp}
%

It is useful to compare the above results with the numerical results of SUS, who adopted Approach II to study the effect of gas pressure on relativistic, maximally rotating SMSs.   Since SUS do not provide results for $\beta = 0$ we extrapolate their numerical values for $\beta > 0$, listed in their Table 1, to $\beta = 0$.  The resulting values for $j_0$ and $\bar M_0$ are listed in our Table \ref{table:crit}.

\begin{table}
\begin{center}
\begin{tabular}{llll}
\hline
\hline
Calculation	& $\delta_j / \beta$ & $\delta_x / \beta$ & $\delta_M^{II} / \beta$ \\
\hline
Paper I; analytical	& - 16.7 & 33.4 &  - 4.7 \\
Paper I; numerical	& - 11.4 & 	22.8 & - 4.6 \\
SUS	& - 13.7 & 27.4	& - 4.6 \\
\hline
\end{tabular}
\end{center}
\caption{Comparison of estimates for $\delta_j$, $\delta_x$ and $\delta_M^{II}$.}
\label{table:coef}
\end{table}

It is evident from Table \ref{table:crit} that the parameters describing the critical configurations show some variation between the different methods.  We did not extract $x_0$ and $j_{\rm min}$ from SUS, but these values agree reasonably well between the analytical and numerical approaches of Paper I.   The largest difference appears for $j_0$, and we note that the value obtained from SUS lies between the two values from Paper I.  All of the above parameters, as well as $\bar M_0$, appear in the expressions (\ref{delta_j_appI}) for $\delta_j$ and expressions (\ref{delta_M_appI})  and (\ref{delta_M_appII}) for $\delta_M$.  Not surprisingly, these perturbations then depend on which background values are adopted.   In Table \ref{table:coef} we list the ratios $\delta_j / \beta$, $\delta_x/ \beta$, and $\delta_M^{II} / \beta$, using the different values of the background parameters (because of the appearance of the $\beta \ln \beta$ term in $\delta_M^{I}$, the ratio $\delta_M^I/\beta$ is not independent of $\beta$; we therefore omitted this quantity in the Table).   For the top two rows in the Table \ref{table:coef} we used the analytical and numerical values from Paper I for all four parameters $x_0$, $j_{\rm min}$, $j_0$ and $\bar M_0$ (i.e.~the values from the corresponding top two rows in Table \ref{table:crit}), while for the bottom entries we used the extra\-polated values from SUS for $j_0$ and $\bar M_0$, but the numerical values of Paper I for $x_0$ and $j_{\rm min}$. 

\begin{figure}
~~~
\begin{center}
\includegraphics[width=3in]{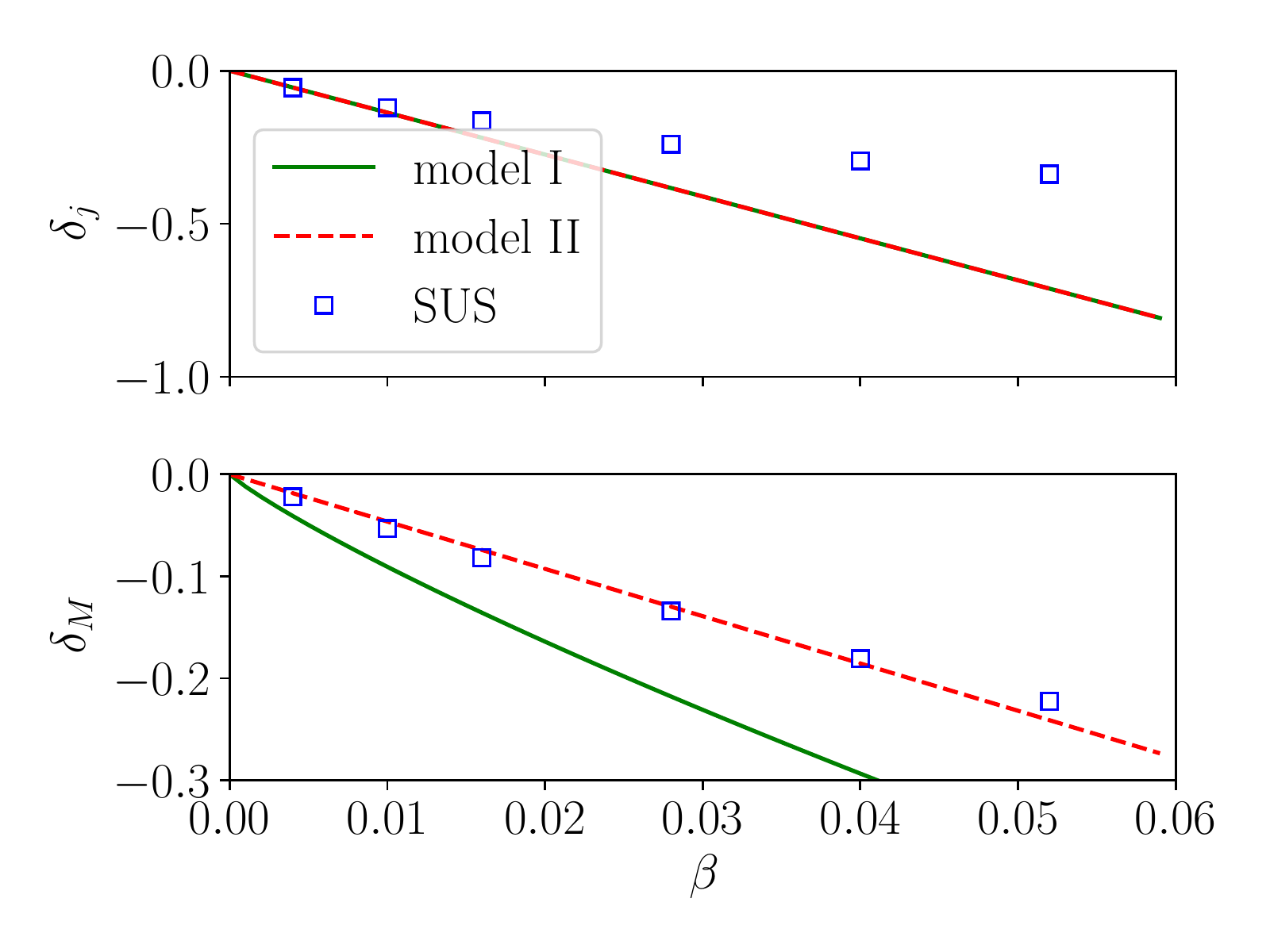}
\end{center}
\caption{Effects of gas pressure on the dimensionless angular momentum $j$ and the rescaled mass $\bar M$ of the critical configuration of rotating SMSs.   We show $\delta_j$ and $\delta_M$, defined in (\ref{delta_j}) and (\ref{delta_M}), as a function of $\beta$.  The solid and dashed lines represent our analytical perturbative expressions according to models I and II, while the squares represent the numerical results of SUS.}
\label{fig:compare}
\end{figure}

For $\delta_j / \beta$ in (\ref{delta_j_appII}) (and similarly for $\delta_x / \beta$) we find variations of several 10\%; these variations are mostly due to differences in the term $2j_0^2 - j_{\rm min}^2$ between the different approaches.  Similar differences in (\ref{delta_M_appII}) tend to cancel each other out, so that the numerical values for $\delta_M^{II}/\beta$ as obtained from different background values agree much better with each other.  We believe that the high-resolution results of SUS provide the most accurate values currently available, and therefore will adopt their parameters in Section \ref{sec:gas:results} below.

We next compare our predictions for $\delta_j$ and $\delta_M$ with the numerical results of SUS.  In order to make this comparison, we compute $\delta_j = (j - j_0)/j_0$ (and similar for $\delta_M$) from the data $j$ listed in Table 1 of SUS, where $j_0$ is the extrapolated value for $\beta = 0$ as listed in our Table \ref{table:crit}.  We also compute values for $\beta$ from $\beta = 6 \,(\Gamma - 4/3)$, where the values of $\Gamma$ are listed in Table 1 of SUS.  In Fig.~\ref{fig:compare} we then graph the numerical values of SUS for $\delta_j$ and $\delta_M$ as squares, and our perturbative results (\ref{delta_j_appI}) as well as (\ref{delta_M_appI}) and (\ref{delta_M_appII}) as solid and dashed lines.  

We first observe again that Approaches I and II make identical predictions for $\delta_j$; they therefore appear as a single line in the top panel of Fig.~\ref{fig:compare}.  Moreover, for small values of $\beta$ the perturbative predictions for $\delta_j$ agree well with the numerical results of SUS.  For larger values of $\beta$ we see increasing deviations.  This is not surprising, however, since, for the larger values of $\beta$ adopted by SUS, our perturbative calculation predicts values of $\delta_j$ of order unity, which is clearly a violation of the condition $|\delta_j| \ll 1$ for a linear treatment (see eq.~(\ref{delta_j})).   Another reason for these deviations at larger $\beta$ may also be related to $T/|W|$, which we assume to remain given by the constant Roche-approximation value (\ref{roche}) independently of $\beta$.  The numerical simulations of SUS show, however, that instead $T/|W|$, which is small, slightly increases with increasing $\beta$.  This means that our approximation underestimates the angular momentum, increasing the deviation in $\delta_j$.

Keeping in mind the subtleties discussed in Section \ref{sec:nonrotSMS:comp}, we also compare our predictions for $\delta_M$ in the lower panel of Fig.~\ref{fig:compare}.  The two different predictions from Approach I and II are included as a solid and a dashed line; as in Fig.~\ref{fig:M_nonrot} for nonrotating SMSs Approach I predicts larger decreases in $\bar M$ than Approach II.   Quite reassuringly, but not surprisingly, Approach II agrees quite well with the results of SUS, who also adopt Approach II.   In fact, for large $\beta$ this agreement is significantly better than for $\delta_j$.  In part, this is because $\delta_M$ is smaller in magnitude than $\delta_j$ for a given value of $\beta$, so that the linearity condition $|\delta_M| \ll 1$ is not violated as severely as the one for $\delta_j$.   We further expect that $\delta_M$ is less affected than $\delta_j$ by our approximation of keeping $T/|W|$ constant, as we discussed above.

%
\subsection{Results for Physical Parameters}
\label{sec:gas:results}
%

Encouraged by the comparisons in the previous Section we now evaluate our perturbative results to compute the changes due to gas pressure in the physical parameters of the critical configuration of a SMS spinning at mass-shedding.  Inserting (\ref{delta_j_appI}) together with (\ref{beta_M}) into (\ref{delta_j}) we find
\begin{equation} \label{JoverM2_pres}
 \left( \frac{J}{M^2} \right)_{\rm crit} \simeq
\left( \frac{J}{M^2} \right)_{\rm crit,0} \left( 1 - 0.12 \left( \frac{M}{10^6 M_{\odot}} \right)^{-1/2} \right)
\end{equation}
and similarly, from (\ref{RoverM1}),
\begin{equation} \label{RoverM_pres}
 \left( \frac{R_p}{M} \right)_{\rm crit} \simeq
\left( \frac{R_p}{M} \right)_{\rm crit,0} \left( 1 - 0.23 \left( \frac{M}{10^6 M_{\odot}} \right)^{-1/2} \right),
\end{equation}
where the unperturbed critical ratios are given in (\ref{JM20}) and (\ref{RoverM0}).  Allowing for gas pressure has a stabilizing effect, permitting SMSs to remain stable to smaller values of $R_p/M$ and higher densities.  For stars with masses $M \lesssim 10^6 M_\odot$ the effect is quite substantial.  In fact, for smaller stars gas pressure dominates the stabilization, as calculated in Section \ref{sec:nonrotSMS}, and rotation should be treated as a perturbation, rather than the other way around as in this Section.

To find the central density we start with (\ref{x}) to find
\begin{equation} 
\rho_{\rm crit} = \frac{x^3}{M^2} \simeq \left( \frac{c^2}{G} \right)^3 \frac{x_0}{M_\odot^2} 
\left( \frac{M_{\odot}}{M} \right)^2 \left(1 + 3 \delta_x \right),
\end{equation}
where we have inserted appropriate powers of $c$ and $G$.  Using (\ref{delta_j_2}) together with (\ref{beta_M}) and evaluating the physical constants we now obtain,
\begin{equation} \label{rho_pres}
\rho_{\rm crit} \simeq \rho_{\rm crit,0}  \left(1 + 0.70 \left( \frac{M}{10^6 M_{\odot}} \right)^{-1/2} \right)
\end{equation}
where $\rho_{\rm crit,0}$ is given by (\ref{rho_crit_0}).  

Eq.~(\ref{rho_pres}) generalizes the earlier result (\ref{rho_crit_phys}): while (\ref{rho_crit_phys}) determines the effect of gas-pressure on the critical density of a nonrotating SMS, eq.~(\ref{rho_pres}) determines the density of a maximally rotating SMS at the onset of instability, allowing for gas pressure as a perturbation.   The two expressions (\ref{rho_crit_phys}) and (\ref{rho_pres}) scale differently with the mass $M$, which is related to the fact that they are based on different expansions.  Eq.~(\ref{rho_pres})  is the result of a perturbation about the nonzero critical density of the maximally spinning SMS.  By contrast, in the absence of rotation, there is no stable configuration without gas pressure, so that, in deriving eq.~(\ref{rho_crit_phys}),  there is no nonzero background critical density about which to expand.  

We note again that the results (\ref{JoverM2_pres}), (\ref{RoverM_pres}) and (\ref{rho_pres}) follow identically from Approaches I and II.

%
\section{Other effects}
\label{sec:others}
%

In this Section we adopt our perturbative approach to estimate the influence of several effects other than gas pressure on the radial stability of uniformly rotating SMSs at the mass-shedding limit.

%
\subsection{Magnetic fields}
%

We now consider the perturbative role of a magnetic field $B$, which we assume to be sufficiently weak so that that the shape of the star remains unchanged and nearly spherical as in Section \ref{sec:rotSMS}.  This assumption requires that $E_{\rm M}/W \ll 1$, where $E_{\rm M}$ is the total magnetic energy.  The energy density of the magnetic field $B$ is
\begin{equation}
\epsilon_{\rm M} = \frac{B^2}{8 \pi}.
\end{equation}
Integrating this energy density of all space gives a result that depends on the topology of the magnetic field.  Typically we find 
\begin{equation} \label{E_M_1}
E_{\rm M} = \kM \Phi_{\rm M}^2 M^{-1/3} \rho_c^{1/3}.
\end{equation}
Here we have expressed the magnetic field in terms of the magnetic flux $\Phi_{\rm M}$ through the matter, which remains constant for a frozen-in magnetic field when we vary the stellar density $\rho_0$.   The constant $\kM$ is a dimensionless quantity that depends on the topology of the magnetic field.  For example, for the model considered by \citet[][p.~242]{Spi78}, which consists of a constant interior field $B$ and a root-mean-square exterior field that falls off as $B (R/r)^3$, we have
\begin{equation}
\kM = \left( \frac{4}{81 \pi^5 C} \right)^{1/3}
\end{equation}
where $C = \rho_c/\langle \rho \rangle \simeq 54.18$ measures the central condensation of an $n=3$ polytope, and where $\Phi_{\rm M} = \pi B R^2$ is the flux through the equatorial plane of the star.   

Using (\ref{x}) and rescaling all dimensional quantities as before, including $\bar \Phi_{\rm M} = K^{-3/2} \Phi_{\rm M}$, we can rewrite (\ref{E_M_1}) in the non-dimensional form
\begin{equation} \label{E_M_2}
\bar E_{\rm M} = \kM \bar \Phi_{\rm M}^2 \bar M^{-1}  x.
\end{equation}
We now add (\ref{E_M_2}) to the energy (\ref{energy1}) to find the total energy 
\begin{align} \label{energy_M}
\bar E  = & ~ k_1 \bar M^{1/3} x + \kM \bar \Phi_{\rm M}^2 \bar M^{-1} x - k_2 \bar M x  \nonumber \\ 
&  + k_3 j^2 \bar M x^2  - k_4 \bar M x^2 - k_5 \bar M x^3.
\end{align}
Setting the first two derivatives to zero yields
\begin{align} \label{denergy_M}
0 = \frac{\partial \bar E}{\partial x}  = & ~ k_1 \bar M^{1/3}  + \kM \bar \Phi_{\rm M}^2 \bar M^{-1}  - k_2 \bar M   \nonumber \\ 
&  + 2 k_3 j^2 \bar M x 
- 2 k_4 \bar M x - 3 k_5 \bar M x^2.
\end{align}
and 
\begin{equation} \label{ddenergy_M}
0 = \frac{\partial^2 \bar E}{\partial x^2} =  2 k_3 j^2 \bar M  
- 2 k_4 \bar M  - 6 k_5 \bar M x.
\end{equation}
Since the magnetic contribution is proportional to $x$, like the internal energy and the Newtonian gravitational potential energy terms, the second derivative of the energy (\ref{energy_M}) with respect to $x$ yields the same equation as (\ref{energy_deriv_2}) for $n=3$.  As a consequence, eqs.~(\ref{x0}) -- (\ref{x2}) and (\ref{j_0}) also remain unchanged.  Solving (\ref{j_0}), again using (\ref{roche}), yields the same values for $j_0$ and $x_0$.  Therefore, the onset of instability occurs at the same values of both $J/M^2$ and $R_p/M$ as in the unperturbed configuration,
\begin{equation}
\left( \frac{J}{M^2} \right)_{\rm crit} =
\left( \frac{J}{M^2} \right)_{\rm crit,0}
\end{equation}
and 
\begin{equation}
\left( \frac{R_p}{M} \right)_{\rm crit} =
\left( \frac{R_p}{M} \right)_{\rm crit,0},
\end{equation}
where the background values are again given in Section \ref{sec:rotSMS:background}.  The sole role of the magnetic perturbation in this approximation is to increase the mass slightly at a given central density, as a result of the magnetic term in (\ref{denergy_M}), but this does not influence the radial instability.  The onset of instability is determined by the competition between rotation and relativistic gravitation, whereas the internal energy, Newtonian gravitational potential energy and magnetic energy establish equilibrium, but are neutral with respect to stability.

We note that in the case of nonrotating SMSs, the magnetic field can lead to a significant increase in the stellar radius of equilibrium configurations \citep[see][as well as Section 7.2 in ST]{OstH68}.

%
\subsection{Dark Matter}
\label{sec:others:DM}
%

We start by considering the effects of a dark-matter (DM) halo.  We treat cold dark matter (CDM) and hot dark matter (HDM) separately, adopting the crude approximations developed in Appendix \ref{sec:E_DM}.  As shown in the appendix, the key parameter that distinguishes these two regimes is $2 \Phi_{\rm SMS}/\varv_\infty^2$, where $\Phi_{\rm SMS}$ is the central potential of the SMS, and $\varv_\infty$ is the speed of the DM particles far from the SMS.   We refer to the limit $2 \Phi_{\rm SMS}/\varv_\infty^2 \gg 1$ as CDM, and the limit $2 \Phi_{\rm SMS}/\varv_\infty^2 \ll 1$ as HDM.   During its evolution the SMS cools and contracts and $\Phi_{\rm SMS}$ increases.  Likewise, during its evolution the Universe expands and $\varv_\infty$ decreases.  Therefore, SMSs may reside in different regimes during different parts of their evolution and different epochs of the Universe.   For example, SMSs arriving at the mass-shedding limit at the current cosmological epoch obey the CDM limit, as we show below, while SMSs arriving at the mass-shedding limit sufficiently early in the Universe, when $\varv_\infty \approx 1$, are in the HDM limit.  

Approximating $\Phi_{\rm SMS}$ as $M/R$ and evaluating this quantity for the critical configuration at the mass-shedding limit, $M/R_p \simeq 1/427$ (see Section \ref{sec:rotSMS:background}), we find
\begin{equation}
\frac{2 \Phi_{\rm SMS}}{\varv_\infty^2} \simeq 4.68 \times 10^3 \left( \frac{\varv_\infty}{300 \,\mbox{km/s}} \right)^{-2}
\end{equation}
where we have normalized to a typical galactic DM speed, appropriate for low redshifts.  In the current cosmological epoch, SMSs therefore reach their critical configuration in the CDM limit, where the above ratio is much greater than unity.

%
\subsubsection{Cold dark matter}
\label{sec:others:DM:cdm}
%

As we argued in Appendix \ref{sec:E_DM:cdm}, the energy contribution due to a cold dark-matter halo can be approximated by the term (\ref{E_CDM_app}),
\begin{equation} \label{E_CDM}
\bar E_{\rm DM} = \kcdm x_{\rm DM}^3 \varv_\infty^{-1} \bar M^{-1/2} x^{-3/2}.
\end{equation}
where $x_{\rm DM} = M^{2/3} (\rho_{\rm DM}^\infty)^{1/3}$.  Including (\ref{E_CDM}) in the energy (\ref{energy1}) and taking the first two derivatives then results in 
 \begin{align} \label{energy_deriv_1_CDM}
0  = & ~ k_1 \bar M^{1/3} - k_2 \bar M
+ 2 k_3 (j^2  - j_{\rm min}^2 ) \bar M x   \\
& - 3 k_5 \bar M x^2 - \frac{3}{2} \kcdm x_{\rm DM}^3 \varv_\infty^{-1} \bar M^{-1/2} x^{-5/2} \nonumber
\end{align}
and
\begin{align} \label{energy_deriv_2_CDM}
0 = & ~  2 k_3 (j^2 - j_{\rm min}^2) \bar M 
- 6 k_5 \bar M x \\ 
&  +  \frac{15}{4} \kcdm x_{\rm DM}^3 \varv_\infty^{-1} \bar M^{-1/2} x^{-7/2}. \nonumber
\end{align}
Dividing (\ref{energy_deriv_2_CDM}) by $\bar M$, multiplying by $x$ (so that, to leading order, the perturbation of $j^2 x = j_0^2 x_0 (1 + 2 \delta_j + \delta_x)$ vanishes according to (\ref{delta_j_2})), and inserting the perturbations (\ref{delta_x}) and (\ref{delta_j}) we obtain
\begin{align} 
&  2 k_3 j_0^2 x_0 - 2 k_3 j_{\rm min}^2 x_0 (1 + \delta_x) 
- 6 k_5 x_0^2 (1 + 2 \delta_x)  = \nonumber \\ 
& ~~~~~ -   \frac{15}{4} \kcdm x_{\rm DM}^3 \varv_\infty^{-1} \bar M_0^{-3/2} x_0^{-5/2}.
\end{align}
Since the DM perturbations scale with $x_{\rm DM}^3$,  the term on the right-hand side is small already, and we can neglect the perturbations of $\bar M$ and $x$ in this term.  The zeroth-order terms yield (\ref{x2}), while the leading-order corrections give
\begin{equation}
( k_3 j_{\rm min}^2 + 6 k_5 x_0 ) \delta_x =  \frac{15}{8} \kcdm x_{\rm DM}^3 \varv_\infty^{-1} \bar M_0^{-3/2} x_0^{-7/2}.
\end{equation}
We now use (\ref{x2}) and (\ref{delta_j_2}) to rewrite the term on the left-hand side to obtain
\begin{equation}
\delta_j = - \frac{15}{16} \frac{\kcdm}{k_3 \bar M_0^{3/2} x_0^{7/2} (2 j_0^2 - j_{\rm min}^2)} \frac{x_{\rm DM}^3}{\varv_\infty}.
\end{equation}
or
\begin{equation}
\delta_j = - 2.5 \times 10^8 \frac{G^3}{c^5} \frac{M^2 \rho_{\rm DM}^\infty}{\varv_\infty},
\end{equation}
where we have inserted the background values of Section \ref{sec:rotSMS:background} as discussed in Section \ref{sec:gas:comp}.   From (\ref{delta_j}) we then have 
\begin{align}
&  \left(\frac{J}{M^2} \right)_{\rm crit} = \left(\frac{J}{M^2} \right)_{\rm crit,0} \times \\
& ~~~\left( 1 - 4.1 \times 10^{-20}  \left( \frac{M}{10^6 M_\odot} \right)^2 \left( \frac{\rho_{\rm DM}^\infty}{10^{-25} \, \mbox{g/cm}^3} \right) 
\left( \frac{\varv_\infty}{300 \, \mbox{km/s} } \right)^{-1} \right), \nonumber
\end{align}
and from (\ref{RoverM1})
\begin{align}
& \left(\frac{R_p}{M} \right)_{\rm crit} = \left(\frac{R_p}{M} \right)_{\rm crit,0} \times \\
& ~~~\left( 1 - 8.2 \times 10^{-20}  \left( \frac{M}{10^6 M_\odot} \right)^2 \left( \frac{\rho_{\rm DM}^\infty}{10^{-25} \, \mbox{g/cm}^3} \right) 
\left( \frac{\varv_\infty}{300 \, \mbox{km/s} } \right)^{-1} \right), \nonumber
\end{align}
where the unperturbed values are given by (\ref{JM20}) and (\ref{RoverM0}).  Here we have adopted values for the DM density and DM particle speeds as they may exist in current DM halos.  Evidently, the effects of such a CDM halo on the stability of rotating SMSs are minute.

%
\subsubsection{Hot dark matter}
\label{sec:others:DM:hdm}
%

In order to account for the effects of a HDM halo on the stability of rotating SMSs we add the expression (\ref{E_HDM_app}),
\begin{equation} \label{E_DM}
\bar E_{\rm DM} = \khdm \bar M^{-2/3} x_{\rm DM}^3 x^{-2},
\end{equation}
 to the energy functional (\ref{energy1}),  where we have again defined $x_{\rm DM} = M^{2/3} (\rho_{\rm DM}^\infty)^{1/3}$.   The first two derivatives of the energy (\ref{energy1}) are then
 \begin{align} \label{energy_deriv_1_DM}
0  = & ~ k_1 \bar M^{1/3} - k_2 \bar M
+ 2 k_3 (j^2  - j_{\rm min}^2 ) \bar M x  \nonumber \\
& - 3 k_5 \bar M x^2 - 2 \khdm x_{\rm DM}^3 \bar M^{-2/3} x^{-3} 
\end{align}
and
\begin{equation} \label{energy_deriv_2_DM}
0 =   2 k_3 (j^2 - j_{\rm min}^2) \bar M
- 6 k_5 \bar M x +6 \khdm x_{\rm DM}^3 \bar M^{-2/3} x^{-4} 
\end{equation}
As in Section \ref{sec:others:DM:cdm} we multiply eq.~(\ref{energy_deriv_2_DM}) with $x$, divide by $\bar M$, perturb about the background solution of Section \ref{sec:rotSMS:background}, and keep the leading-order correction terms to obtain
\begin{equation}
0 = - 2 k_3 j_{\rm min}^2 x_0 \delta_x - 12 k_5 x_0^2 \delta_x + 6 \khdm x_{\rm DM}^3  \bar M_0^{-5/3} x_0^{-3}
\end{equation}
(where again all corrections scale with $x_{\rm DM}^3$, so that this term is a small quantity already.)  Using eq.~(\ref{delta_j_2}) as well as (\ref{x2}) we may rewrite this expression as
\begin{equation} \label{delta_j_DM1}
\delta_j = - \frac{3}{2} \frac{\khdm \bar M_0^{-5/3}}{k_3 (2 j_0^2 - j_{\rm min}^2)} \frac{x_{\rm DM}^3}{x_0^4}.
\end{equation}
or
\begin{equation} \label{delta_j_DM2}
\delta_j = - \frac{3}{2} \frac{\khdm \bar M_0^{-5/3}}{k_3 (2 j_0^2 - j_{\rm min}^2) x_0^4 } \left( \frac{G}{c^2} \right)^3 M^2 \rho_{\rm DM}.
\end{equation}
Evaluating the background quantities as discussed in Section \ref{sec:gas:comp} and inserting into (\ref{delta_j}) we find
\begin{equation} \label{delta_j_CDM}
 \left( \frac{J}{M^2} \right)_{\rm crit} \simeq
 \left( \frac{J}{M^2} \right)_{\rm crit,0} \left( 1 - 0.038 \left( \frac{M}{10^6 M_\odot} \right)^2 \left( \frac{\rho_{\rm DM}}{10^{-5} \,\mbox{g/cm}^{3}} \right) \right),
\end{equation}
where we followed \citet{McLF96} and \citet{Bis98} in adopting a DM density of $10^{-5} \, \mbox{g/cm}^3$, appropriate for some epoch in the early Universe.  Inserting (\ref{delta_j_DM2}) into (\ref{RoverM1}) we also find
\begin{equation} \label{RoverM_CDM}
 \left( \frac{R_p}{M} \right)_{\rm crit} \simeq
\left( \frac{R_p}{M} \right)_{\rm crit,0} \left( 1 - 0.077 \left( \frac{M}{10^6 M_\odot} \right)^2 
\left( \frac{\rho_{\rm DM}}{10^{-5} \, \mbox{g/cm}^3} \right) \right).
\end{equation}
Even for a large value of the DM density, the effect of the DM on the critical configuration of uniformly rotating SMSs is relatively small unless $M > 10^6 M_{\odot}$, which is consistent with the findings of \citet{McLF96} and \citet{Bis98}.

%
\subsection{Dark energy}
\label{sec:others:CC}
%

The effects of a cosmological constant can be estimated very similarly to that of constant DM density distribution in the HDM regime.  The Newtonian limit of Einstein's equations, including a cosmological constant $\Lambda$, is 
\begin{equation}
\nabla^2 \Phi = 4 \pi \rho - \Lambda.
\end{equation}
We can therefore find the effects of the cosmological constant simply by replacing the constant value of $\rho_{\rm HDM}$ in the equations of Section \ref{sec:others:DM:hdm} with $- \Lambda/(4 \pi)$.   Making this replacement in (\ref{delta_j_DM2}) and inserting physical constants we find
\begin{equation}  \label{delta_j_CC}
\delta_j = \frac{3}{8 \pi} \frac{\khdm}{k_3} \frac{1}{\bar M_0^{5/3} (2 j_0^2 - j_{\rm min}^2) x_0^4 } \left( \frac{G}{c^2} \right)^2 M^2 \Lambda.
\end{equation}
Evaluating the coefficients using the background quantities of Section \ref{sec:rotSMS:background} and inserting into (\ref{delta_j}) we find
\begin{equation}
\left( \frac{J}{M^2} \right)_{\rm crit} \simeq
\left( \frac{J}{M^2} \right)_{\rm crit,0} 
\left( 1 +  4.5 \times 10^{-26} \left( \frac{M}{10^6 M_\odot} \right)^2 \left( \frac{\Lambda}{\Lambda_{\rm SM}} \right) \right),
\end{equation}
where we have adopted
\begin{equation} \label{Lambda_SM}
\Lambda_{\rm SM} = 1.11 \times 10^{-56} \, \mbox{cm}^{-2} 
\end{equation}
for the cosmological constant in the standard model \citep{Planck15}.   Inserting (\ref{delta_j_CC}) into (\ref{RoverM1}) we similarly obtain 
\begin{equation} \label{RoverM_CC}
\left( \frac{R_p}{M} \right)_{\rm crit} \simeq
\left( \frac{R_p}{M} \right)_{\rm crit,0} \left( 1 + 0.91 \times 10^{-25} \left( \frac{M}{10^6 M_\odot} \right)^2 \left( \frac{\Lambda}{\Lambda_{\rm SM}} \right) \right).
\end{equation}
Evidently, the effects of the dark energy on the stability of typical SMSs are entirely negligible.

%
\section{Summary}
\label{sec:summary}
%

In this paper we extend a calculation initiated in Paper I, where we determined the critical configuration at the onset of collapse of a uniformly rotating SMS, supported by pure radiation pressure and spinning at the mass-shedding limit.  We found that this critical configuration is characterized by a unique set of the non-dimensional parameters $R_p/M$ and $J/M^2$.  

In this case, the subsequent collapse of the critical configuration follows a universal evolutionary track and results in a spinning black hole with $M_{\rm BH}/M \approx 0.9$ and $J_{\rm BH}/M_{\rm BH}^2 \approx 0.7$ \citep{ShiS02,ShaS02}, surrounded by a disk with $M_{\rm disk}/M \approx 0.1$.  Moreover, if the initial star is threaded by a weak, toroidal magnetic field, then the black hole-disk system launches a jet, and the electromagnetic Poynting luminosity transported by the jet resides in a narrow range of $L_{\rm EM} \approx 10^{52 \pm 1} \, \mbox{ergs/s}$, independent of mass, once the system settles into quasi-steady accretion \citep[see][]{SunPRS17,Sha17}).   The universal evolutionary track also leads to the emission of a universal gravitational signal \citep{ShiSUU16,SunPRS17}, which may serve as a ``standard siren" for future space-based gravitational wave detectors.

In this paper we explore the domain of validity of this universality by considering a number of different physical effects that might perturb this rotating critical configuration.  In particular, we study the effects of gas pressure, magnetic fields, dark matter and dark energy, and find that they perturb the critical parameters $J/M^2$ and $R/M$ according to
\begin{align} \label{JoverM2_summary}
\left( \frac{J}{M^2} \right)_{\rm crit} & \simeq  
\left( \frac{J}{M^2} \right)_{\rm crit,0} \Bigg( 1 - 0.12 \left( \frac{M}{10^6 M_{\odot}} \right)^{-1/2} \\
& - 4.1 \times 10^{-20} \left( \frac{M}{10^6 M_\odot} \right)^2 \left( \frac{\rho_{\rm DM}}{10^{-25} \, \mbox{g/cm}^3} \right) 
	\left( \frac{\varv_\infty}{300 \, \mbox{km/s}} \right)^{-1} \nonumber \\
& + 4.5 \times 10^{-26} \left( \frac{M}{10^6 M_\odot} \right)^2 \left( \frac{\Lambda}{\Lambda_{\rm SM}} \right) \Bigg),
\nonumber
\end{align}
and
\begin{align} \label{RoverM_summary}
\left( \frac{R_p}{M} \right)_{\rm crit} &  \simeq 
\left( \frac{R_p}{M} \right)_{\rm crit,0} \Bigg( 1 - 0.23 \left( \frac{M}{10^6 M_{\odot}} \right)^{-1/2} \\
& - 8.2 \times 10^{-20} \left( \frac{M}{10^6 M_\odot} \right)^2 \left( \frac{\rho_{\rm DM}}{10^{-25} \, \mbox{g/cm}^3} \right) 
	\left( \frac{\varv_\infty}{300 \, \mbox{km/s}} \right)^{-1} \nonumber \\
& + 0.91 \times 10^{-25} \left( \frac{M}{10^6 M_\odot} \right)^2 \left( \frac{\Lambda}{\Lambda_{\rm SM}} \right) \Bigg),
\nonumber
\end{align}
where the unperturbed values are given by eqs.~(\ref{JM20}) and (\ref{RoverM0}).

Gas pressure is the most significant perturbation for the most realistic astrophysical scenarios, and becomes important for SMSs with masses of $M \lesssim 10^6 M_\odot$.  Its effects are accounted for by the top lines in (\ref{JoverM2_summary}) and (\ref{RoverM_summary}).  We model these effects using two different approaches (see Section \ref{sec:thermo:appI} and \ref{sec:thermo:appII}),  which we carefully compare and calibrate (see Section \ref{sec:nonrotSMS}).  Both approaches lead to identical results for the leading-order effects on $J/M^2$ and $R/M$, which also agree well with the numerical results of SUS in the linear regime.   Gas pressure has a stabilizing effect on rotating SMSs, it scales with $M^{-1/2}$ and becomes important for $M \lesssim 10^6 M_\odot$.  

It turns out that magnetic fields have an effect on the mass of the critical configuration of maximally rotating SMSs, but not on the dimensionless ratios $J/M^2$ and $R_p/M$ -- the effect of magnetic fields is therefore absent in eqs.~(\ref{JoverM2_summary}) and (\ref{RoverM_summary}).

We approximate the effects of a DM halo adopting two opposite limiting cases, which we refer to as CDM and HDM.  Rotating SMSs that reach the onset of instability in the current cosmological epoch are in the CDM regime, which we include in the middle lines of (\ref{JoverM2_summary}) and (\ref{RoverM_summary}).  The effects of DM are stabilizing, but are minute and can be neglected for most situations.  For SMSs forming in the early universe, the HDM regime may apply; the result, rescaled for a higher DM density as it might have applied in the early Universe, can be found in eqs.~(\ref{delta_j_CDM}) and (\ref{RoverM_CDM}).

Finally, the effects of dark energy are included in the last lines of (\ref{JoverM2_summary}) and (\ref{RoverM_summary}); these effects are destabilizing but even smaller than those for of a DM halo.  

In general relativity, SMSs supported by angular momentum and pure radiation pressure can be stabilized against collapse only for angular momenta $j$ that are greater than $j_{\rm min}$ defined in (\ref{j_min}), which is close to the mass-shedding value (\ref{JM20}).  Thus for our perturbative calculations, which take configurations stabilized by rotation and radiation pressure as background models, slowly rotating stars as described by, e.g., \citet{MaeM00,YooKK15,HaeWKHW18a}, are not suitable.  SMSs that are rotating well below the mass-shedding limit could be modeled as configurations in which gas pressure plays the dominant stabilizing role and rotation is a small perturbation -- but these are not the configurations that we focus on in this paper.

\section*{Acknowledgments}

SPB and ARL gratefully acknowledge support through undergraduate research fellowships at Bowdoin College.   This work was supported in part by NSF grants PHY-1402780 and PHY-1707526 to Bowdoin College, as well as NSF grants PHY-1602536 and PHY-1662211 and NASA grants NNX13AH44G and 80NSSC17K0070 to the University of Illinois at Urbana-Champaign (UIUC).  TWB would like to thank UIUC for extending their hospitality during a visit.

\begin{appendix}

%
\section{The dark matter contribution to the stellar energy}
\label{sec:E_DM}
%

In this appendix we derive approximations for the contribution of a dark-matter (DM) halo on the energy of a SMS, considering the opposite extremes of cold and hot dark matter.  We imagine that the SMS is surrounded by a much larger region filled with DM particles that are collisionless, monoenergetic and moving isotropically.  Far from the SMS the DM has a uniform density $\rho_{\rm DM}^\infty$ and the particle speed is $\varv_\infty \ll 1$.   A simple phase-space distribution function describing this DM is 
\begin{equation}
f = f(E) = \rho_{\rm DM}^{\infty} \frac{\delta(E - E_\infty)}{4 \pi (2 E_\infty)^{1/2}},
\end{equation}
where the asymptotic energy per unit mass is $E_\infty = \varv_\infty^2/2$ (see (ST.14.2.16)).    Integrating over the distribution  function we find for the density
\begin{equation} \label{rho_DM}
\rho_{\rm DM} = \rho_{\rm DM}^{\infty} \left( 1 - \frac{2 \Phi_{\rm SMS}}{v_{\infty}^2} \right)^{1/2}
\end{equation}
(compare (ST.14.2.22)).  

Strictly speaking, the above distribution function holds only for stationary configurations.  It is also a very good approximation for the quasistationary situation encountered when we consider small oscillations of the equilibrium configuration.  Justification of this approximation is provided by the following argument \citep{McLF96} which suggests that the DM density is at most very weakly affected by the oscillations of the SMS.   

Consider a dark-matter particle with zero speed $\varv_\infty$ at a large separation from the SMS.  Once it has reached the surface of the SMS, it will travel at the escape speed $v \simeq (GM/R)^{1/2}$, and will therefore transverse the star in a time of approximately $\tau_{\rm cross} \sim R/v \sim (R^3/(GM))^{1/2} \sim  (G \rho)^{-1/2} \sim \tau_{\rm ff}$, where $\tau_{\rm ff}$ is the free-fall time.  The crossing time $\tau_{\rm cross}$ will be even shorter for particles with $\varv_\infty >0$ at large separation, so that $\tau_{\rm cross} \lesssim \tau_{\rm ff}$ in general.   For stars with structures similar to $n = 3$ polytropes, the fundamental radial oscillation period $\tau_{\rm osc}$ is larger than than the free-fall time, which establishes the inequality $\tau_{\rm cross} < \tau_{\rm osc}$.   In fact, for stars near the critical configuration, the oscillation frequency approaches zero, so that $\tau_{\rm osc}$ becomes infinite.   Assuming spherical symmetry, dark-matter particles are unaffected by oscillations of the SMS while they are outside of the star, because the exterior potential will not change.  Since $\Phi_{\rm SMS}$ varies on the time scale $\tau_{\rm osc}$, which is longer than the crossing time $\tau_{\rm cross}$, the quasi-stationary approximation applies, and hence we may adopt eq.~(\ref{rho_DM}).

We now consider two opposite extreme limits, namely $2 \Phi_{\rm SMS}/\varv_\infty^2 \gg 1$ and $2 \Phi_{\rm SMS}/\varv_\infty^2 \ll 1$, where $\Phi_{\rm SMS}$ is evaluated inside the SMS.  The former applies in the limit when the speed of the dark-matter particles is much less than the escape speed from the SMS, which we we refer to as CDM.  Conversely, the latter applies when the dark-matter particles are much faster than the escape speed, which we refer to as HDM.  We treat these two limits separately in the following two Sections.  At large distances from the SMS we always have $\Phi_{\rm SMS} \rightarrow 0$, so that, from (\ref{rho_DM}), we always have $\rho_{\rm DM} \rightarrow \rho_{\rm DM}^\infty$ in both limits, as expected.

%
\subsection{Cold dark matter}
\label{sec:E_DM:cdm}
%

In a region where $2 \Phi_{\rm SMS}/\varv_\infty^2 \gg 1$ we may approximate (\ref{rho_DM}) as 
\begin{equation} 
\rho_{\rm CDM} \simeq \rho_{\rm DM}^\infty \left( - \frac{2 \Phi_{\rm SMS}}{v_{\infty}^2} \right)^{1/2}.
\end{equation}
We now invoke a further approximation and argue that, given the large central condensation of an $n=3$ polytrope, we may crudely approximate the potential of the SMS as that of a point mass, so that
\begin{equation} \label{rho_CDM}
\rho_{\rm CDM} \simeq \rho_{\rm DM}^\infty \left( \frac{2 M}{r v_{\infty}^2} \right)^{1/2}.
\end{equation}
The Newtonian potential caused by the dark-matter distribution can then be found from the Poisson equation
\begin{equation} \label{poisson}
\nabla^2 \Phi_{\rm CDM} = \frac{1}{r^2} \frac{d}{dr} \left( r^2 \frac{d \Phi_{\rm CDM}}{dr} \right) = 4 \pi \rho_{\rm CDM}.
\end{equation}
Inserting (\ref{rho_CDM}) and integrating twice we obtain
\begin{equation} \label{Phi_CDM}
\Phi_{\rm CDM} = 2^{1/2} \frac{16 \pi}{15} \frac{\rho_{\rm DM}^\infty M^{1/2}}{\varv_\infty} r^{3/2} + C,
\end{equation}
where $C$ is a constant of integration, and where we have set a second constant of integration to zero in order to make $\Phi_{\rm CDM}$ regular at the origin.  We can now find the contribution of this potential to the energy of the SMS from
\begin{align} \label{CDM_integral_1}
E_{\rm CDM} & =  \int_0^M \Phi_{\rm CDM} dm = 4 \pi \int_0^R \Phi_{\rm CDM} \rho r^2 dr \nonumber \\
	& =  2^{1/2} \frac{64 \pi^2}{15} \frac{\rho_{\rm DM}^\infty M^{1/2}}{\varv_\infty} \int_0^R \rho r^{7/2} dr. 
\end{align}
Here the integration is carried out over the SMS, we have assumed spherical symmetry, and we have omitted the constant $C$ in (\ref{Phi_CDM}) since it would lead to a term that is independent the density, and which would therefore drop out when we take derivatives of the energy with respect to the density.  To leading order, the structure of the SMS is that of an $n=3$ polytrope, so that we can rewrite the last integral in (\ref{CDM_integral_1}) using Lane-Emden variables $\rho = \rho_c \theta^3$ and $r = a_{\rm LE} \xi$ with $a_{\rm LE} = K^{1/2} \rho_c^{-1/3} / \pi^{1/2}$,
\begin{align} \label{CDM_integral_2}
E_{\rm CDM} 
	& = 2^{1/2} \frac{64 }{15 \pi^{1/4}} \frac{\rho_{\rm DM}^\infty M^{1/2}}{\varv_\infty} \rho_c a_{\rm LE}^{9/2}  
	 \int_0^{\xi_3} \theta^3 \xi^{7/2} d\xi \\
	 & = 2^{1/2} \frac{64 }{15 \pi^{1/4}} \frac{\rho_{\rm DM}^\infty M^{1/2}}{\varv_\infty} \rho_c^{-1/2} K^{9/4}  
	 \int_0^{\xi_3} \theta^3 \xi^{7/2} d\xi.  \nonumber
\end{align}
We now define
\begin{equation} \label{k_CDM}
\kcdm \equiv 2^{1/2} \frac{64 }{15 \pi^{1/4}} \int_0^{\xi_3} \theta^3 \xi^{7/2} d\xi = 30.0193
\end{equation}
and use the 
dimensional rescaling $\bar E = K^{-3/2} E$ (and similar for $M$) together with (\ref{x}) and
\begin{equation}
x_{\rm DM} = M^{2/3} (\rho_{\rm DM}^\infty)^{1/3}
\end{equation}
to obtained the rescaled energy 
\begin{equation} \label{E_CDM_app}
\bar E_{\rm CDM} = \kcdm x_{\rm DM}^3 \varv_\infty^{-1} \bar M^{-1/2} x^{-3/2}.
\end{equation}
We may now account for the effects of cold DM by including this term in the energy (\ref{energy1}).

%
\subsection{Hot dark matter}
\label{sec:E_DM:hdm}
%

In the opposite limit, $2 \Phi_{\rm SMS}/\varv_\infty^2 \ll 1$, the dark-matter density is unaffected by the SMS and takes the constant value
\begin{equation} \label{rho_HDM}
\rho_{\rm HDM} \simeq \rho_{\rm DM}^\infty,
\end{equation}
which is the regime considered by \citet{McLF96} and \citet{Bis98}.  Solving the Poisson equation (\ref{poisson}) for the potential $\Phi_{\rm HDM}$ we obtain
\begin{equation}
\Phi_{\rm HDM} = \frac{2 \pi}{3} \rho_{\rm DM}^{\infty} r^2,
\end{equation}
where we have omitted two constants of integration for the same reasons as in Section \ref{sec:E_DM:cdm}.  The contribution of this potential to the energy of the SMS is then
\begin{align} \label{DM_integral}
E_{\rm HDM} & =  \int_0^M \Phi_{\rm HDM} dm = 4 \pi \int_0^R \Phi_{\rm HDM} \rho r^2 dr \nonumber \\
& =  \frac{8 \pi^2}{3} \rho_{\rm DM}^\infty \int_0^R \rho r^4 dr.
\end{align}
As in Section \ref{sec:E_DM:cdm} we use Lane-Emden variables for an $n=3$ polytrope to rewrite this integral as 
\begin{align} \label{DM_integral2}
E_{\rm HDM} & =   \frac{8 \pi^2}{3} \rho_{\rm DM}^\infty \rho_c a_{\rm LE}^5 \int_0^{\xi_3} \theta^3 \xi^4 d \xi \nonumber \\
& = \frac{8}{3 \pi^{1/2}} K^{5/2} \rho_{\rm DM}^\infty \rho_c^{-2/3} \int_0^{\xi_3} \theta^3 \xi^4 d \xi. 
\end{align}
We now define
\begin{equation} \label{k_HDM}
\khdm = \frac{8}{3 \pi^{1/2}} \int_0^{\xi_3} \theta^3 \xi^4 d \xi = 16.3262.
\end{equation}
and adopt the same rescaling as in Section \ref{sec:E_DM:cdm} to find the rescaled energy
\begin{equation} \label{E_HDM_app}
\bar E_{\rm HDM} = \khdm \bar M^{-2/3} x_{\rm DM}^3 x^{-2}.
\end{equation}
In order to account for a hot dark-energy halo, we therefore add this term to the energy (\ref{energy1}). 

\end{appendix}

\bibliographystyle{mnras}

\end{document}